\documentclass[fleqn,usenatbib]{mnras}

\usepackage{newtxtext,newtxmath}

\usepackage[T1]{fontenc}

\DeclareRobustCommand{\VAN}[3]{#2}
\let\VANthebibliography\thebibliography
\def\thebibliography{\DeclareRobustCommand{\VAN}[3]{##3}\VANthebibliography}

\usepackage{graphicx}	
\usepackage{amsmath}	
\usepackage{listings}
\usepackage{siunitx}
\usepackage{tikz}
\usepackage{subcaption}
\usepackage{hyperref}

\newcommand{\angstrom}{\text{\normalfont\AA}}

\newcommand{\ba}{\[\begin{aligned}}
\newcommand{\ea}{\end{aligned}\]}

\makeatletter
\g@addto@macro\bfseries{\boldmath}
\makeatother

\DeclareSIUnit \parsec {pc}
\DeclareSIUnit \h {h}
\DeclareSIUnit\year{yr}

\title[The Physics of Lyman Continuum Escape]{The Physics of Indirect Estimators of Lyman Continuum Escape and their Application to High-Redshift JWST Galaxies}

\author[Choustikov, N., et al.]{Nicholas Choustikov$^{1}$\thanks{nicholas.choustikov@physics.ox.ac.uk},
Harley Katz$^{1}$,
Aayush Saxena$^{1,2}$,
Alex J. Cameron$^{1}$,
Julien Devriendt$^{1}$, \newauthor
Adrianne Slyz$^{1}$,
Joki Rosdahl$^{3}$, 
Jeremy Blaizot$^{3}$,
and Leo Michel-Dansac$^{3}$
\\
$^{1}$Sub-department of Astrophysics, University of Oxford, Keble Road, Oxford OX1 3RH, United Kingdom\\
$^{2}$Department of Physics and Astronomy, University College London, Gower Street, London WC1E 6BT, United Kingdom\\
$^{3}$CNRS, Centre de Recherche Astrophysique de Lyon UMR5574, Univ Lyon, Univ Lyon1, Ens de Lyon, F-69230 Saint-Genis-Laval, France
}

\date{Accepted XXX. Received YYY; in original form ZZZ}

\pubyear{2023}

\begin{document}
\label{firstpage}
\pagerange{\pageref{firstpage}--\pageref{lastpage}}
\maketitle

\begin{abstract}
Reliable indirect diagnostics of LyC photon escape from galaxies are required to understand which sources were the dominant contributors to reionization. While multiple LyC escape fraction ($f_{\rm esc}$) indicators have been proposed to trace favourable conditions for LyC leakage from the interstellar medium of low-redshift ``analog'' galaxies, it remains unclear whether these are applicable at high redshifts where LyC emission cannot be directly observed. Using a library of 14,120 mock spectra of star-forming galaxies with redshifts $4.64 \leq z \leq 10$ from the SPHINX$^{20}$ cosmological radiation hydrodynamics simulation, we develop a framework for the physics that leads to high $f_{\rm esc}$. We investigate LyC leakage from our galaxies based on the criteria that successful LyC escape diagnostics must \textit{i)} track a high specific star formation rate, \textit{ii)} be sensitive to stellar population age in the range $3.5-10$~Myr representing the times when supernova first explode to when LyC production significantly drops, and \textit{iii)} include a proxy for neutral gas content and gas density in the interstellar medium. ${\rm O}_{32}$, $\Sigma_{\rm SFR}$, M$_{\rm UV}$, and H$\beta$ equivalent width select for one or fewer of our criteria, rendering them either necessary but insufficient or generally poor diagnostics. In contrast, UV slope ($\beta$), and ${\rm E(B-V)}$ match two or more of our criteria, rendering them good $f_{\rm esc}$ diagnostics (albeit with significant scatter). Using our library, we build a quantitative model for predicting $f_{\rm esc}$ based on direct observables. When applied to bright $z > 6$ Ly$\alpha$ emitters observed with JWST, we find that the majority of them have $f_{\rm esc} \lesssim 10\%$.
\end{abstract}




\section{Introduction}
\label{sec:introduction}
Various astrophysical \citep[e.g.][]{Keating2020,Kulkarni_2019,Becker_2021} and cosmological \citep{Planck_2018} probes indicate that the Universe had transitioned from a neutral to an ionized state by the redshift interval $5\lesssim z\lesssim6$. However, the onset of reionization and the sources responsible for it, as well as the neutral fraction evolution during the transition remain uncertain. Both upper ($z \sim 20$) and lower ($z \sim 15$) limits on the onset of cosmic dawn are provided by the Cosmic Microwave Background \citep[e.g.][]{Heinrich2021} and star formation histories (SFHs) of high-redshift galaxies \citep[e.g.][]{Laporte2021}. Although often model-dependent, neutral fraction evolution constraints can be derived from the decreasing prevalence of Ly$\alpha$ emitters \citep[e.g.][]{Pentericci2011,Stark2010,Mason2018,Jones:2023,Goovaerts:2023}, the damping wings of high-redshift quasars \citep[e.g.][]{Durovcikova2020,Davies2018,Greig2019}, and the opacity of the Ly$\alpha$ forest \citep[e.g.][]{Fan2006,Bosman2022}. However, only limited observational constraints exist on the sources responsible for reionization. 

Understanding the sources of reionization is of key importance. The topology of reionization is strongly affected by the source model, which not only impacts the shape and amplitude of the 21~cm signal, \citep[e.g.][]{Zaldarriaga2004,McQuinn2007,Kulkarni2017}, but also controls which dwarf galaxies and filaments are regulated by radiation feedback \citep{Katz2020-reion}. Furthermore, the temperature the IGM reaches during reionization is sensitive to the spectral energy distribution (SED) of the sources responsible. 

Some debate exists about whether these sources are active galactic nuclei (AGN) or star-forming galaxies. AGN-dominated models of late reionization have been suggested to explain observational constraints such as the low optical depth to Thompson scattering \citep[e.g.][]{Madau:2015, Chardin:2017, Grissom:2014, Torres-Alba:2020}. Furthermore, recent observations have identified numerous AGN-candidates at high redshift \citep{Larson:2023, Maiolino:2023, Fujimoto:2023}, suggesting that there may be more of them than previously expected \citep{Greene:2023}. In contrast, models \citep{Faucher-Giguere:2020} based on the local X-ray background \citep{Parsa:2018}, the fact that helium reionization was completed significantly later at $z \sim 3$ \citep[e.g.][]{Furlanetto:2008} and an apparent drop in the AGN luminosity function with increasing redshift \citep{Kulkarni_2019} (although this is based on observations taken before the launch of JWST) all indicate that AGN had a sub-dominant effect on the reionization history of the universe. This agrees with other results indicating that AGN only became a dominant source of ionizing photons at redshifts of $z \sim 4$ \citep[e.g.][]{Dayal:2020, Trebitsch_2021}.  

As a result, Star-forming galaxies are often considered the primary candidates for providing the bulk of the LyC photons for reionization \citep[e.g.][]{Robertson2015, Livermore:2017, Naidu_2020}. Within the galaxy population, it is generally assumed that reionization was driven by dwarf galaxies due to the steep observed faint-end slopes of the high-redshift UV luminosity function \citep[e.g.][]{Bouwens2022,Harikane2023}. However this latter assumption is highly dependent on the amount of LyC photons that are produced per unit star formation (or UV luminosity) as well as the  fraction of LyC photons that leak ($f_{\rm esc}$) as a function of mass (or UV luminosity).

The former can be estimated from stellar population synthesis models \citep[e.g.][]{Leither1999,Stanway_2018}. While uncertainties in the ionizing photon production rate exist due to systematic differences between stellar population models (e.g. binaries, rotation, IMF, etc.), the escape fraction is far less constrained. This is due to the fact that it emerges from complex, highly non-linear physics (describing e.g. the state of the ISM) and, more importantly, it cannot be directly detected during the epoch of reionization due to the increasingly neutral IGM \citep{Inoue_2014}. For these reasons, constraints on $f_{\rm esc}$ are derived indirectly, for example, by observing samples of low-redshift LyC leaking galaxies that are considered ``analogs'' of those that form during the epoch of reionization \citep[e.g.][]{Izotov_2018b,Flury_2022a}, by directly modelling LyC leakage with cosmological radiation hydrodynamics simulations \citep[e.g.][]{Paardekooper2015,Rosdahl_2018}, or by correlating galaxies with Ly$\alpha$ forest transmission \citep[e.g.][]{Kakiichi2018}.

The number of observational measurements of $f_{\rm esc}$ are rapidly growing. LyC photons are directly detectable with space-based facilities, such as at $z\sim0.3$ with the cosmic origins spectrograph on HST \citep{Green_2012,Leitherer_2016} or at even higher redshifts with AstroSat \citep{Saha2020}. Ground-based and space-based observatories have pushed the redshift frontier of LyC measurements to $z\gtrsim3$ \citep[e.g.][]{Vanzella2010,Steidel_2018,Fletcher2019,Saxena_2022}. The Low Redshift Lyman Continuum Survey \citep[LzLCS,][]{Flury_2022a,Flury_2022b} in particular has significantly increased the total number of low-redshift galaxies with detected LyC emission. However, it remains unclear whether these ``analogs'' are truly representative of the high-redshift galaxy population \citep{Katz_2022b,Brinchmann2022,Schaerer_2022a,Katz_2023}. Moreover, it is not always clear how to generalize results from observations of individual objects or surveys with complex selection functions to the general population of high-redshift galaxies. 

Numerical simulations provide a complementary framework to understand the physics of LyC leakage. However, self-consistently modelling the production and transfer of LyC photons through a resolved, multiphase ISM remains a technically challenging problem. Nevertheless, simulations of individual or a few high-redshift galaxies are commonplace \citep[e.g.][]{Kimm_2017, Trebitsch_2017,Kimm_2014,Ma_2020}, though these suffer from similar generalizability arguments. Larger volume or full-box cosmological radiative transfer simulations that resolve the ISM for thousands of galaxies are now becoming technically feasible \citep[e.g.][]{Xu_2016,Rosdahl_2018}. SPHINX$^{20}$ \citep{Rosdahl_2022} represents such an effort where the connection between $f_{\rm esc}$ and various galaxy properties (such as stellar and halo mass, UV luminosity, star formation rate (SFR), specific star formation rate (sSFR), metallicity, etc.) can be studied across a sample of $>10,000$ galaxies per snapshot between $4.64\leq z\lesssim15$. 

Unfortunately, the connection between simulations and observations remains limited. With the exception of Ly$\alpha$ \citep[e.g.][]{Verhamme2015,Kakiichi_2021,Kimm2019,Kimm2022,Maji2022}, simulations tend to focus on how $f_{\rm esc}$ varies with ``unobservable'' quantities such as halo mass. Efforts have been made to mock observations \citep[e.g.][]{Mauerhofer2021,Barrow2020,Katz_2020,Zackrisson_2017} and infer LyC escape; however, these remain a small minority. In contrast, observational-based studies often focus on indirect diagnostics of $f_{\rm esc}$. For example, such diagnostics include high $[\mathrm{OIII}]\lambda5007/[\mathrm{OII}]\lambda\lambda3726,3728$ (O$_{32}$) ratios \citep[e.g.][]{Jaskot_2013,Nakajima_2014,Izotov_2018}, Ly$\alpha$ peak separation \citep[e.g.][]{Verhamme_2017, Izotov_2020} or Ly$\alpha$ equivalent width \citep{Steidel_2018,Pahl_2021}, [Mg~{\small II}]~$\lambda\lambda2796,2804$ doublet ratios \citep{Chisholm_2020, Katz_2022b}, strong [C~{\small IV}]~$\lambda\lambda1548,1550$ emission \citep{Schaerer_2022,Saxena_2022}, UV slope \citep{Chisholm_2022}, and S~{\small II} deficits \citep{Wang_2021}.

In this work, we address the applicability of various indirect, observationally-developed, diagnostics of LyC escape on a statistical sample of simulated high-redshift galaxies that are likely to be observable with JWST. We develop a physically motivated model for the conditions that need to be met to ensure both high $f_{\rm esc}$ and a simultaneous significant production of LyC photons. Using mock observations from Version~1 of the SPHINX Public Data Release (SPDR1), we discuss how various indirect diagnostics fit within our framework to elucidate the physics of why a leakage indicator is successful (or not). We intend our results to be immediately applicable to the large samples of JWST galaxies currently being observed at $z>6$ \citep[e.g.][]{Curtis-Lake2022,Finkelstein2022,Treu2022,Matthee2022b}.

This work is organized as follows. In Section \ref{sec:methods} we outline the numerical methods behind our new SPHINX$^{20}$ data-set. Section \ref{sec:framework} presents and tests our new generalised framework for identifying LyC leaking galaxies. In Section \ref{sec:discussion} we use our framework to contextualise and explain known diagnostics and use our data-set to predict escape fractions from JWST spectra. Finally, caveats are given in Section \ref{sec:caveat} and we conclude in Section \ref{sec:conclusion}.

\section{Numerical Simulations}
\label{sec:methods}
Due to the observational challenges of both detecting LyC radiation and being limited to individual lines of sight, we employ state-of-the-art numerical simulations to understand both the physics driving LyC leakage and the observational signatures of a high escape fraction. More specifically, we use the SPHINX$^{20}$ cosmological radiation hydrodynamics simulation \citep{Rosdahl_2022}. This simulation is ideal for our purposes because the volume ($20^3$~cMpc$^3$) is large enough to sample a wide diversity of galaxy stellar masses ($10^4 M_{\odot} - 10^{10} M_{\odot}$) and properties, while the spatial and mass resolution\footnote{The maximum level of refinement corresponds to a physical scale of $\sim10$~pc at $z=6$ while the star and dark matter particles have masses of $400\ {\rm M_{\odot}}$ and $2\times10^5\ {\rm M_{\odot}}$, respectively.} allow us to simultaneously resolve many of the low mass haloes that may be contributing to reionization as well as much of the multiphase ISM in the simulated galaxies. Full details of the simulation are described in \citep{Rosdahl_2018,Rosdahl_2022}.

\begin{figure}
	\includegraphics{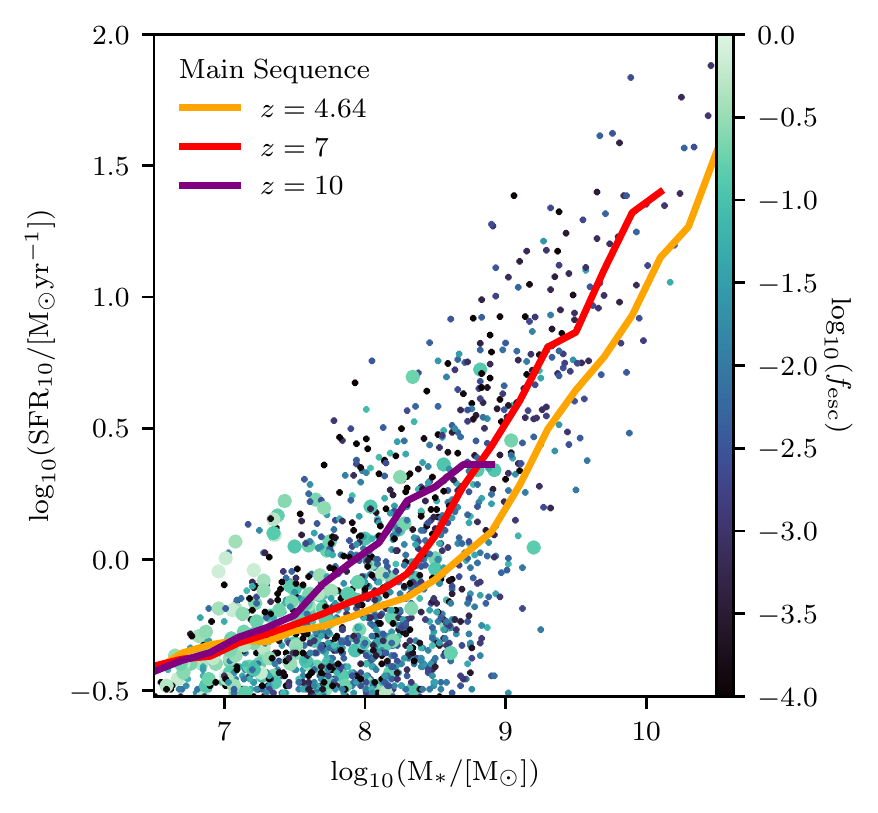}
   \caption{Star formation rate (averaged over $\SI{10}{\mega\year}$) as a function of stellar mass. Each data-point is coloured by the LyC escape fraction, with those with fractions above 10\% enlarged. This shows that LyC leakers tend to fall above the main sequence fitted for galaxies at $z = \{4.64, 7, 10\}$ (orange, red, and purple respectively), representing starburst periods.}
    \label{fig:MS}
\end{figure}

We select a subset of observable galaxies from SPHINX$^{20}$ in the snapshots at $z = 10,9,8,7,6,5,4.64$, that have $\SI{10}{\mega\year}$-averaged SFRs $\geq0.3~\mathrm{M}_{\odot}\SI{}{\per\year}$. The SFR threshold is designed to mimic a flux limited survey and allows us to study a select group of epoch of reionization galaxies from SPHINX$^{20}$ that are most likely to be detectable by deep JWST observations\footnote{We use intrinsic UV magnitudes and best-fit cosmological parameters from \cite{Planck_2018} to find that the dimmest galaxies at $z=4.64$ and $10$ have intrinsic apparent magnitudes of $30.9$ and $32.3$ respectively. Current JWST NIRCAM surveys therefore have sufficient depth to image the majority ($87\%$) of these galaxies \citep[e.g.][]{Eisenstein:2023,Casey:2023, Bagley:2023}. Comparisons between mock SPHINX$^{20}$ and JADES photometry are shown in Figure 15 of \citet{spdrv1}, showing that this is a good comparative sample.}. The total sample contains 1,412 galaxies and in Figure~\ref{fig:MS} we show the SFR versus stellar mass for our entire sample, along with the main sequences for the subset of galaxies at $z = \{4.64, 7, 10\}$ in orange, red, and purple respectively. Due to the SFR threshold, the main sequence follows a curve, calculated here using a moving average. The fact that LyC leakers fall around the main sequence at their redshift has been discussed at length before. Interested readers are directed to \cite{Katz:2023b}. Hereafter, data displayed will contain galaxies from each of these redshift bins.

For each galaxy we compute a total intrinsic SED by summing emission from the star particles and gas cells. Stellar emission follows the BPASS v2.2.1 SED \citep{Stanway_2018}, and is computed based on the mass, age, and metallicity of each star particle. Line emission from each gas cell is computed using an updated version of the method presented in \cite{Katz_2022f}. For each gas cell, we calculate the total ionizing flux. For cells that host star particles, we sum the contributions from all star particles within the cell\footnote{This assumes that the local star particles dominate the ionizing photon budget of the gas cell. This assumption may fail when a cell has neighbours with much higher star formation efficiencies.}. For gas cells without star particles, we use the ionizing fluxes directly from the RT solver in the simulation. 

Like all cosmological simulations, SPHINX$^{20}$ is limited by spatial resolution, which in this case can pose a problem for the few radiation fronts from individual star particles which are not completely resolved. In the case of such un-resolved Stromgren spheres, we need to be careful as the temperature of the gas cell ends up being an average between those of the ionised and neutral phases. This leads to a lower effective gas temperature in the H~{\small II} region, which primarily impacts collisionally excited lines (although in many cases not their ratios) and to a lesser-degree, recombination lines, while having almost no affect on IR fine structure lines or resonant scatter. In order to mitigate this issue, we apply a sub-resolution model. First, we identify all cells that host star particles where the Stromgren radius ($R_{\rm S}$) is unresolved (i.e. $R_{\rm S}<\Delta x/2$). For cells without an unresolved Stromgren sphere, we proceed as normal and line emissivities are calculated with {\small CLOUDY} v17.03 \citep{Ferland_2017} as these cells are unaffected by the issues discussed above\footnote{This statement holds true except when a thin ionization front is moving through the cell, but this case is very difficult to identify with a Eulerian code.}. We tabulate a grid of one-zone slab models varying the gas density, metallicity, ionization parameter, and electron fraction. All models are iterated to convergence and the shape of the SED varies with metallicity but is assumed to have an age of $10$~Myr\footnote{Note that the normalization of the SED in these cells matters significantly more than the exact spectral shape. We have opted for 10~Myr as this roughly corresponds to the time when significant SN could have disrupted the gas in the host cell which allows radiation to leak to neighbouring cells.}. To calculate the emission from unresolved Stromgren spheres, we run a second grid of {\small CLOUDY} models assuming a spherical geometry, varying stellar age, metallicity, gas density, and total ionizing luminosity. Stellar age is computed as the ionizing luminosity-weighted average stellar age in each gas cell. Finally, we use the line emissivities as calculated by the appropriate spherical {\small CLOUDY} model, which now implicitly uses the corrected temperature. By definition, radiation from the unresolved Stromgren sphere does not leak to surrounding gas cells, meaning that this fix does not affect them. For a further discussion, the reader is directed to \cite{spdrv1}. The total intrinsic emission line luminosity of a galaxy is then the sum of all gas cells within the virial radius.

To better compare with observations we also account for dust scattering and absorption. Following \cite{Katz_2022b}, we use {\small RASCAS} \citep{Michel_Dansac_2020} and the effective SMC dust model from \cite{Laursen_2009} to attenuate the SEDs and line emission. Since the attenuation depends on viewing angle, we employ the peeling algorithm \citep{Yusuf_Zadeh_1984,Zheng_2002,Dijkstra_2017} to compute the full dust-attenuated SED for ten different uniformly distributed directions. Hence our full data set consists of 14,120 simulated spectra, though we will in some cases discuss angle-averaged versions of these quantities. We then use these spectra to extract dust-attenuated observables, including line luminosities, equivalent widths, UV spectral index ($\beta$, by fitting to $f_{\lambda} \propto \lambda^{\beta}$ around 1500~\angstrom), UV attenuation given by the Balmer decrement (${\rm E(B-V)}$), UV magnitude (M$_{\rm UV}$), and the star formation rate surface density ($\Sigma_{\rm SFR}$). In order to compare like-for-like, $\Sigma_{\rm SFR}$ is measured by a SFR converted from H$\beta$ luminosity \citep{Kennicutt1998} as well as the dust attenuated half-light radius at 1500~\angstrom.

\begin{figure}
	\includegraphics{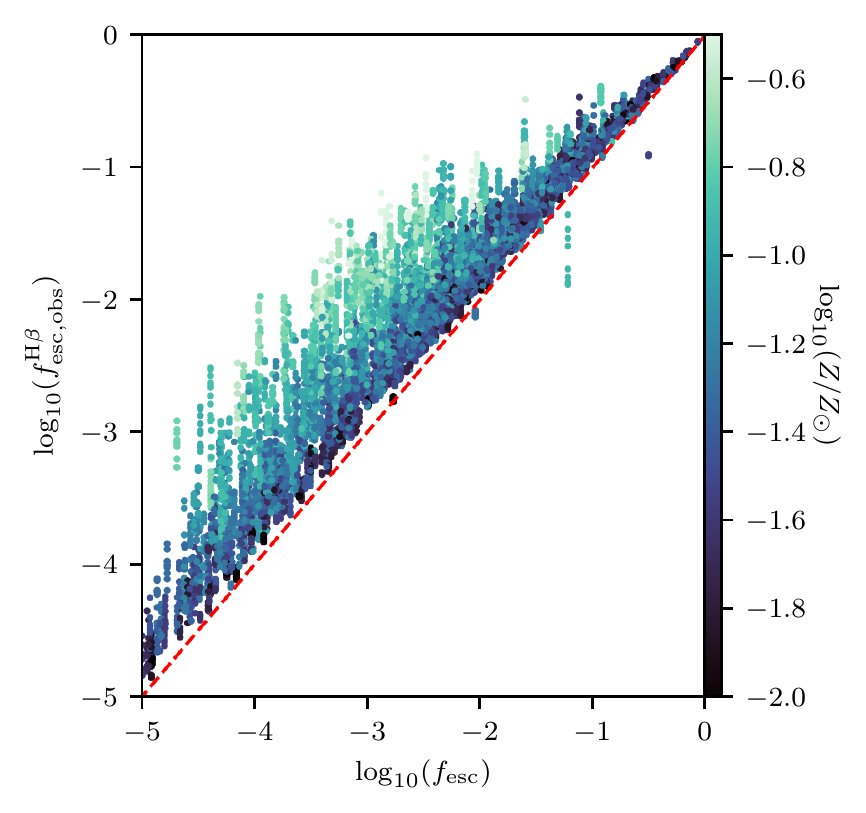}
    \caption{Line-of-sight H$\beta$-derived LyC escape fractions as a function of true LyC escape fraction for SPHINX$^{20}$ galaxies, coloured by the gas metallicity. Metal-rich systems tend to over-predict $f_{\rm esc}$ as compared to the true value.}
    \label{fig:fesc_fesc_Hb}
\end{figure}

We have also post-processed the simulation to measure $f_{\rm esc}$ for all galaxies in our sample. Angle-averaged LyC escape fractions are calculated with {\small RASCAS} \citep{Michel_Dansac_2020} by ray-tracing LyC emission from star particles (see \citealt{Rosdahl_2022}). While this value can be used to measure the instantaneous contribution of each halo to reionization, it cannot be easily measured with observations due to both the wavelength coverage and the deep exposure times needed. For this reason, we measure a second value of $f_{\rm esc}$ for each line of sight using the H$\beta$ luminosity such that \citep{Izotov_2016}:
\begin{equation}
\label{eq:fesc_obs}
    f_{\rm esc,obs}^{\mathrm{H}\beta}=\frac{f_{\rm esc}L_{\rm LyC}}{L_{\rm H\beta}/4.86\times10^{-13}+f_{\rm esc}L_{\rm LyC}}.
\end{equation}
This allows us to make a more fair comparison with observational surveys such as LzLCS \citep{Flury_2022b} where such a method is used. A comparison between the two $f_{\rm esc}$ measurements is shown in Figure~\ref{fig:fesc_fesc_Hb}. While there is a strong correlation between the two quantities for the highest values of $f_{\rm esc}$, in general, the H$\beta$ method tends to over-predict the true value. This is due to the fact that the normalization constant of $4.86\times10^{-13}$ which is commonly used in the literature \citep[e.g.][]{Matthee2022b} is not fully representative of the value in our simulation\footnote{By fitting intrinsic H$\beta$ to total ionizing flux for our galaxies we find a value of $\SI{4.05e-13}{}$. However, in order to best compare to observational methods we use the theoretical value. \citep[e.g.][]{Schaerer_2003}}. This bias will not qualitatively impact the general trends we find between $f_{\rm esc}$ and various observational quantities. Henceforth, any reference to $f_{\rm esc}$ invokes the angle-averaged value derived by {\small RASCAS}, while $f_{\rm esc}^{\mathrm{H}\beta}$ refers to the value derived by Equation \ref{eq:fesc_obs}.

\section{A Generalised Framework for LyC Leakage}
\label{sec:framework}

\begin{figure*}
	\includegraphics{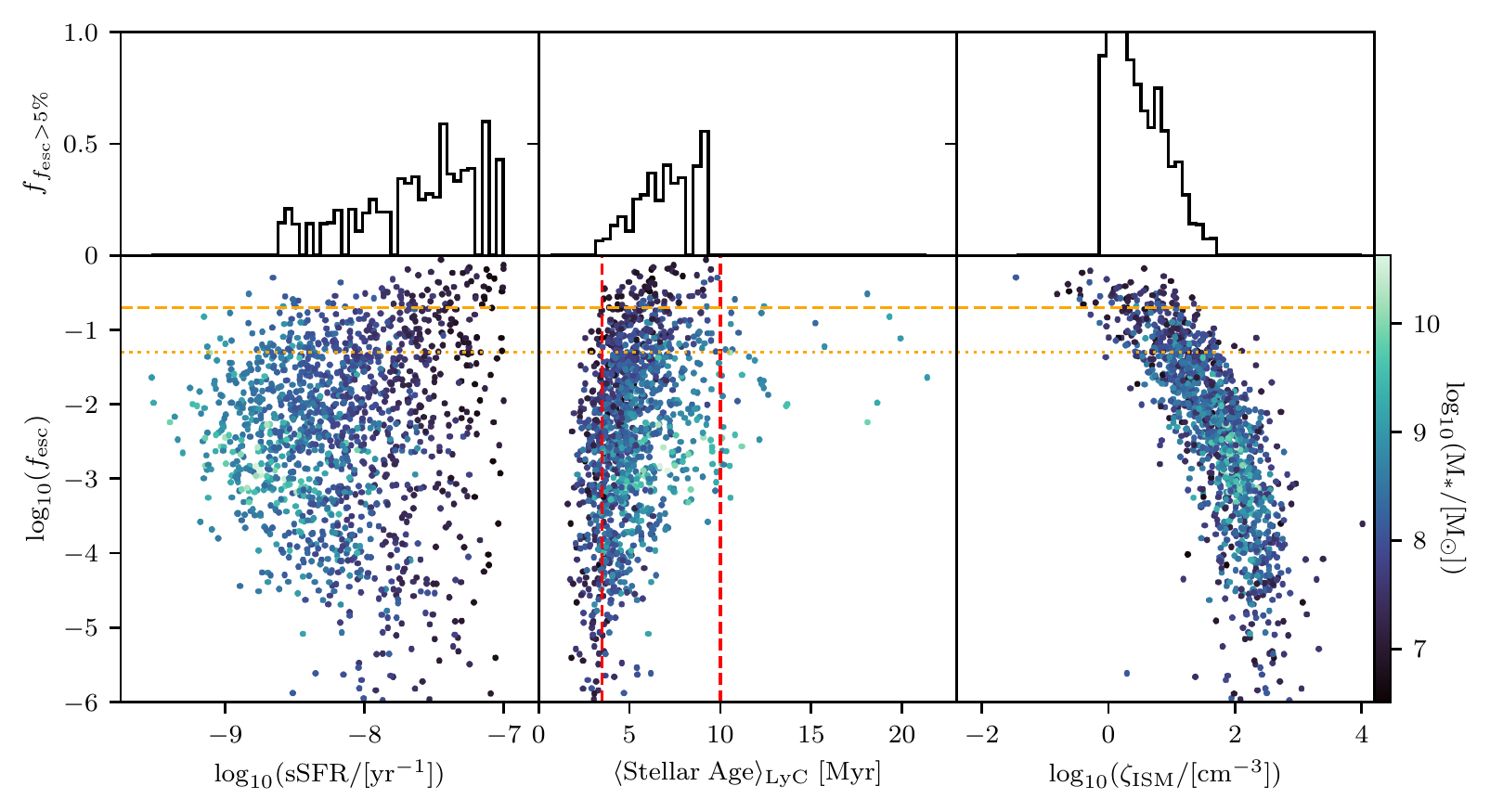}
   \caption{LyC escape fraction as a function of specific star formation rate (left), ionizing flux-weighted mean stellar age (center) and angle-averaged composite neutral gas attenuation parameter $\zeta_{\rm ISM} = \mathrm{E(B-V)\times\langle n_{\rm H}\rangle_{[\rm OII]}}$ (right), all coloured by stellar mass. Escape fractions of 20\% and 5\% are marked with dashed and dotted horizontal orange lines, respectively. The top panels show a histogram of the fraction of galaxies with $f_{\rm esc} > 5\%$ in each bin with more than 5 such galaxies. Systems with the highest escape fractions tend to have high sSFR, fall within $\SI{3.5}{\mega\year}$ and $\SI{10}{\mega\year}$ (indicated by red lines), low neutral gas attenuation parameter, and have low stellar masses. However, each of these requirements is a necessary but insufficient diagnostic for LyC leakers.}
    \label{fig:three_crit}
\end{figure*}
Numerous diagnostics for identifying galaxies with high $f_{\mathrm{esc}}$ have been suggested in the literature (see Section \ref{sec:introduction}); however the vast majority have been shown to be ``necessary but insufficient'' conditions for LyC leakage \citep{Flury_2022b}. We therefore aim to provide a more general framework describing the conditions necessary for galaxies to both produce and leak a significant amount of ionizing LyC radiation.

In order for a galaxy to meaningfully contribute to reionization it must simultaneously be producing copious amounts of ionizing photons and a fraction of those photons must be able to leak into the low-density IGM where the recombination timescale is longer than the Hubble time. While simulations have yet to quantitatively agree on the average escape fractions of galaxies as a function of various galaxy properties (e.g. mass, \citealt{Ma_2020, Rosdahl_2022}), qualitatively they nearly all find that $f_{\mathrm{esc}}$ is a feedback-related quantity \citep[e.g.][]{Trebitsch_2017}. Energetic feedback from stars (e.g. supernovae (SN)) disrupt the ISM, creating ionized channels through which photons leak into the IGM. Thus, based on such a scenario we propose that in order to produce a significant amount of LyC leakage there needs to be:
\begin{itemize}
    \item A strong burst of star formation such that a significant quantity of ionizing photons are produced.

    \item Either a significant reduction in the neutral gas content of the ISM such that photons can leak relatively isotropically (density-bounded case) or a creation of ionized channels (ionization-bounded case). This is achieved through photoionization and mechanical feedback.

    \item A timescale synchronization such that stars continue producing significant amounts of LyC photons after feedback has disrupted the ISM. 
\end{itemize}

Therefore, we argue that a good diagnostic to identify LyC leakers should simultaneously encapsulate:
\begin{enumerate}
    \item High sSFR so that significant numbers of ionizing photons are produced and feedback has a chance of disrupting the ISM. 

    \item A stellar population age $\gtrsim3.5$~Myr such that SN have had time to explode but $\lesssim10$~Myr so that the LyC production rate remains high.

    \item A proxy for neutral gas content and the ionization state of the ISM to identify when feedback has efficiently coupled to gas in order to create ionized channels or disrupt/ionize the entire medium.
\end{enumerate}

In Figure~\ref{fig:three_crit} we demonstrate that each of these conditions alone are insufficient to select a sample of only LyC leaking galaxies. In the left panel we show $f_{\rm esc}$ versus sSFR (coloured by total stellar mass) and while many of the leakers have high sSFR, in general there is no trend between the two quantities \citep[see also][]{Rosdahl_2022}. Similar behaviour is also seen in observations \citep[e.g.][]{Flury_2022b,Saxena_2022}. The centre panel shows $f_{\rm esc}$ versus mean stellar age weighted by LyC luminosity (to highlight the age of stars which contribute to LyC flux). In order to reach an escape fraction of 20\% (above the dashed orange line), the age must be $\gtrsim3.5$~Myr as indicated by the left-most vertical red line. However, selecting by age alone clearly results in significant contamination.  Our final criteria is a representative proxy for the state of the ISM. It is difficult to find a proxy that only contains information about the ISM and not the SFR or age as these are highly coupled. Nevertheless, for demonstrative purposes we have chosen a composite parameter $\zeta_{\rm ISM} = {\rm E(B-V)\times\langle n_H\rangle_{[OII]}/[cm^{-3}]}$ where we multiply the angle-averaged UV attenuation ${\rm E(B-V)}$, (which has already been shown by \cite{Saldana-Lopez2022} to empirically correlate with $f_{\rm esc}$) by the gas density weighted by the intrinsic [O~{\small II}]~$\lambda\lambda$3727 luminosity\footnote{Note that we find qualitatively similar results when replacing density with the intrinsic ratio of [O\ {\small II}]~$\lambda3729$/[O\ {\small II}]~$\lambda3726$.}. While there is clearly a strong trend and all galaxies with $f_{\rm esc}>20\%$ have a value $\log_{10}(\zeta_{\rm ISM})\lesssim1.3$, there are many non-leakers with such low values as well. These findings are reinforced by the histogram above each subplot showing the fraction of galaxies with escape fractions higher than 5\% (i.e. above the dotted orange line) in each bin containing more than five such leakers.

\begin{figure}
    \includegraphics{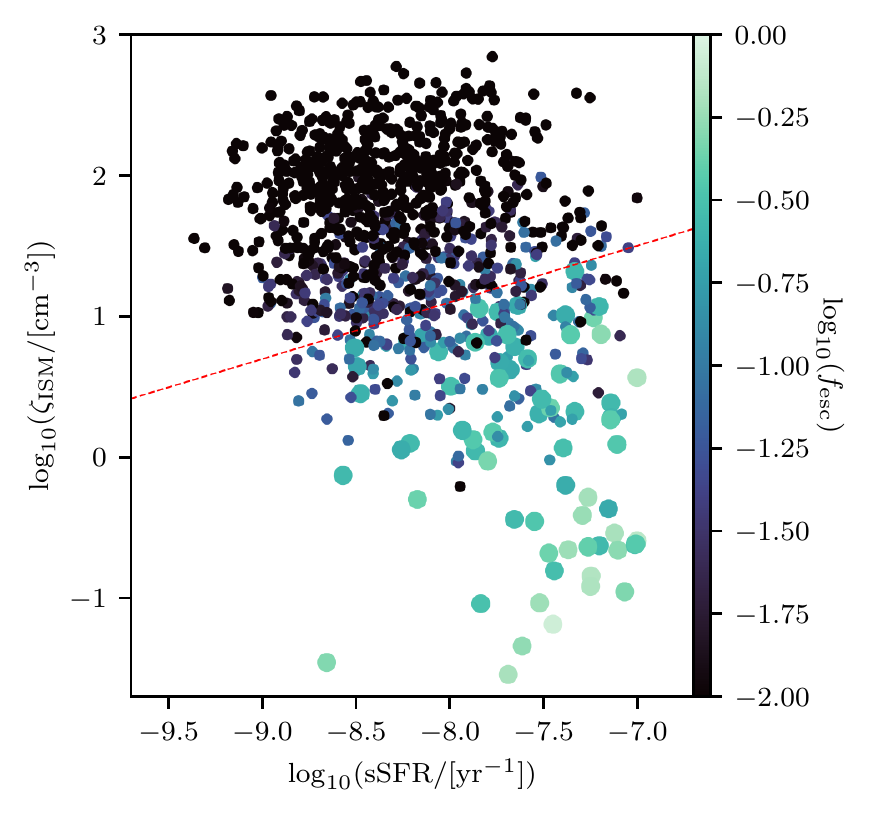}
    \includegraphics{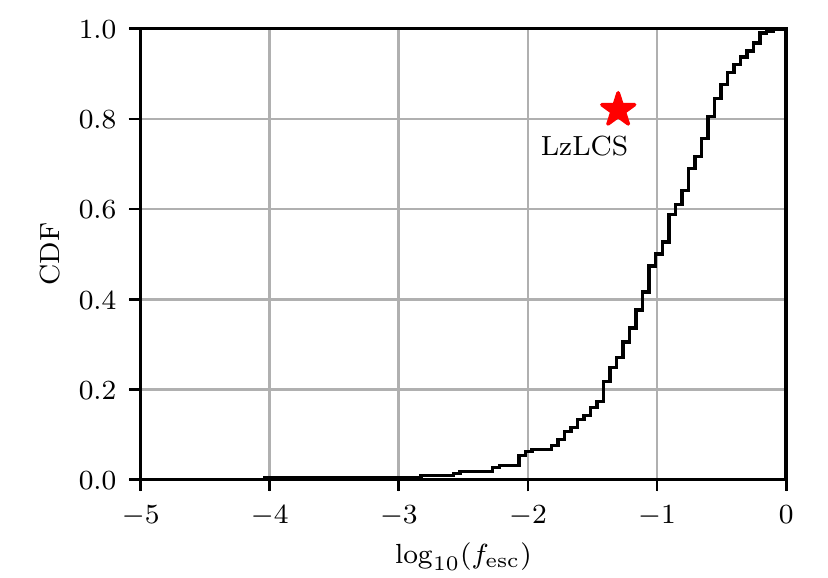}
    \caption{(Top) angle-averaged neutral gas parameter $\zeta_{\rm ISM}$ as a function of specific star formation rate for galaxies with mean stellar ages between $\SI{3.5}{\mega\year}$ and $\SI{10}{\mega\year}$ coloured by LyC escape fraction. Systems with $f_{\rm esc}>20\%$ are enlarged. A selection criteria (given by Equation \ref{eq:theoretical_cut}) to produce a sample highly enriched with leakers, 74\% of which have $f_{\rm esc} > 5\%$ is shown as the red line. (Bottom) Cumulative distribution for escape fractions for galaxies that satisfy our selection criteria. For comparison, the red star shows that $82\%$ of galaxies selected to be part of LzLCS have $f_{\rm esc} < 5\%$, while 26\% of SPHINX$^{20}$ galaxies within these criteria have $f_{\rm esc} < 5\%$.}
    \label{fig:three_crit_sel}
\end{figure}

It is clear that none of these three conditions alone are sufficient for identifying a contamination-free sample of LyC leakers. However, together, they provide a robust framework for identifying LyC leakers. In Figure~\ref{fig:three_crit_sel} we plot angle-averaged $\zeta_{\rm ISM}$ versus sSFR for SPHINX$^{20}$ galaxies with $3.5\leq{\rm \langle Stellar\ Age\rangle_{LyC}\ /[Myr]}<10$. The galaxies with high $f_{\rm esc}$ are biased towards having high sSFR and low $\zeta_{\rm ISM}$. By selecting galaxies with
\begin{equation}
    \log_{10}(\zeta_{\rm ISM}/[\mathrm{cm}^{-3}]) < 0.4\times\log_{10}({\rm sSFR}/[
    \mathrm{yr}^{-1}])+4.3, \label{eq:theoretical_cut}
\end{equation}
we generate a sample based on angle-averaged quantities that is biased towards having high $f_{\rm esc}$. In the bottom panel of Figure~\ref{fig:three_crit_sel} we show a cumulative distribution function for the escape fractions of galaxies that satisfy our selection criteria. This consists of 227 galaxies (16\% of our sample), $\sim74\%$ of which have $f_{\rm esc}>5\%$. These leakers account for $\sim 65\%$ of all such leakers in our sample. This can be contrasted with LzLCS where only 12/66 galaxies (18\%) have $f_{\rm esc}>5\%$. While the efficiency of our selection is not yet perfect, our framework provides a theoretically motivated model with minimal contamination from galaxies with low $f_{\rm esc}$.

\begin{figure*}
    \begin{subfigure}{6.97in}
        \includegraphics{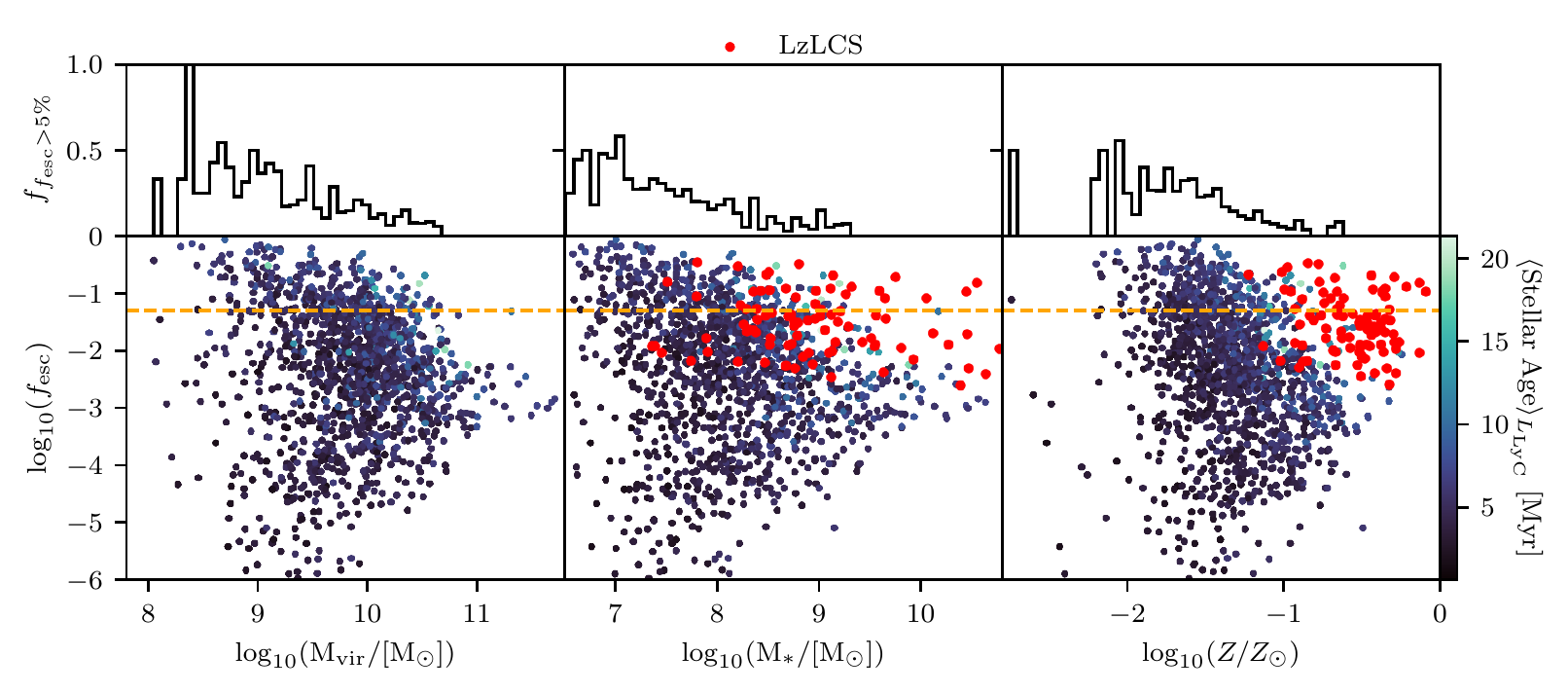}
    \end{subfigure}
    \begin{subfigure}{6.97in}
        \includegraphics{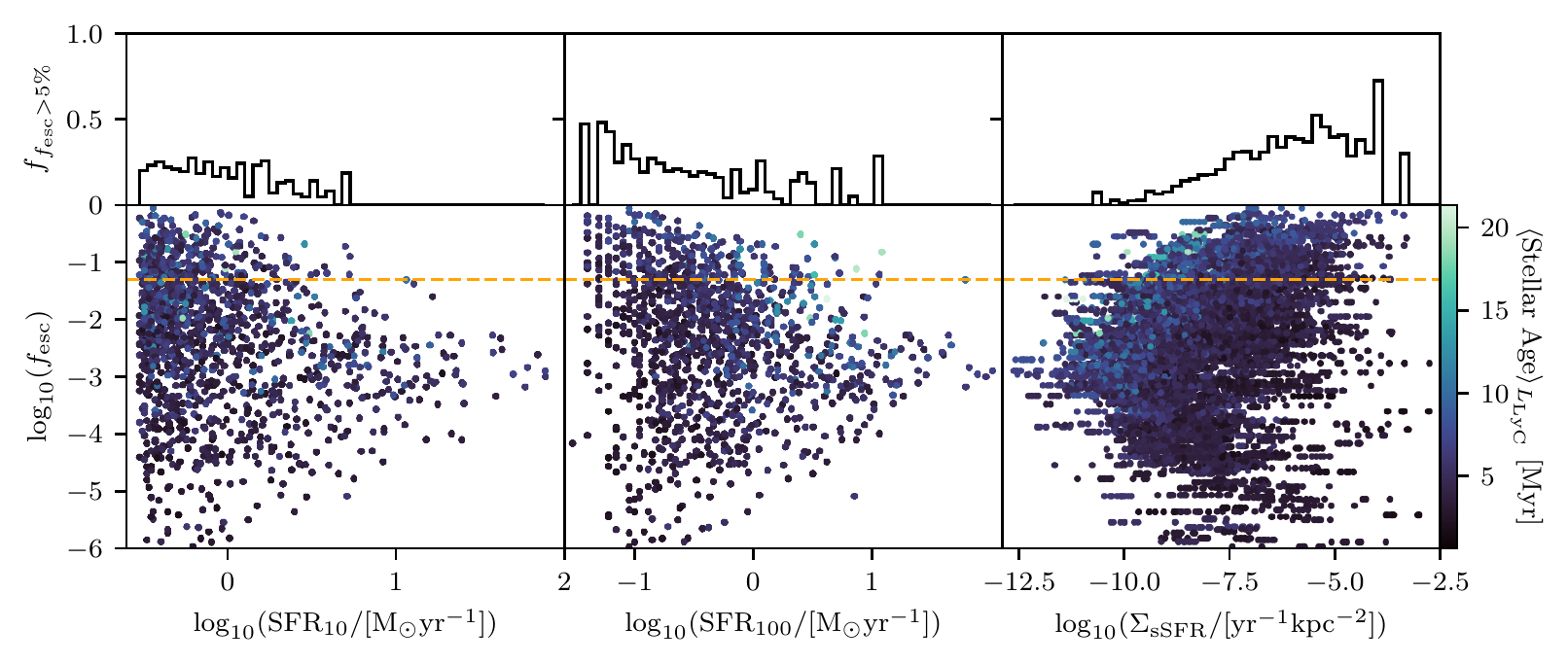}
    \end{subfigure} 
   \caption{LyC escape fraction as a function of halo virial mass (top-left), stellar mass (top-center), gas metallicity (top-right), $\SI{10}{\mega\year}$ averaged SFR (bottom-first), $\SI{100}{\mega\year}$ averaged SFR (bottom-center), and sSFR surface density (bottom-right) coloured by the LyC luminosity-weighted mean stellar age. Where available, LzLCS data is shown in red for comparison. For each quantity, a histogram is given for the number density of galaxies with $f_{\rm esc} > 5\%$ in each bin (shown by the dashed orange line).}
    \label{fig:mega_three_criteria}
\end{figure*}

Despite the effectiveness of such a selection method, it is important to note that thus far this discussion has focused on galaxy properties that are not directly observable. Nevertheless, for completeness, in Figure~\ref{fig:mega_three_criteria} we plot $f_{\rm esc}$ versus halo mass, stellar mass, metallicity, 10~Myr-averaged SFR, 100~Myr-averaged SFR, and the sSFR surface density. As a sub-sample of SPHINX$^{20}$ galaxies, the data set presented here displays many of the trends described in \cite{Rosdahl_2022} relating to which galaxy properties correlate with $f_{\rm esc}$. However, there are subtle differences because we have selected only star-forming galaxies. There is a minor tendency for $f_{\rm esc}$ to decrease with increasing halo mass and stellar mass, consistent with \cite{Razoumov2010,Kimm_2014,Paardekooper2015,Xu_2016,Ma_2020,Rosdahl_2022}. Because of our cut in SFR, we do not sample a downturn in $f_{\rm esc}$ that is seen at lower stellar and halo masses in some simulations \citep{Ma_2020,Rosdahl_2022}. 

It is well established observationally that gas-phase metallicity scales with stellar mass \citep[e.g.][]{Lequeux1979,Tremonti2004}. Such a trend holds in SPHINX$^{20}$ and for this reason, the trend of $f_{\rm esc}$ with metallicity is also similar to that of $f_{\rm esc}$ with stellar mass. Likewise, there exists a star formation main sequence that correlates stellar mass and SFR \citep[e.g.][]{Brinchmann2004,Salim_2007}, as shown for our sample in Figure~\ref{fig:MS}. Hence we see similar behaviour between SFR$_{10}$ and $f_{\rm esc}$ as for stellar mass and $f_{\rm esc}$. Though we see the same broad behaviour for SFR$_{100}$ (albeit with more scatter), low longer-term SFR seems to better select for LyC leakers. Furthermore, the prevalence of star formation burstiness (using ${\rm SFR}_{10}/{\rm SFR}_{100}$) in LyC leakers has already been discussed in depth using the same data-set. For this, the reader is directed to \cite{Katz:2023b}.

While few trends exist between these fundamental galaxy properties and $f_{\rm esc}$, the histogram of each sub-Figure shows that there are certain regions of parameter space where one is more likely to find LyC leakers. For example, there are a higher fractions of leakers at low virial masses, stellar masses, and gas metallicities, indicating that the conditions needed for leakage are more often accessible in these environments. Similar behaviour is also observed in LzLCS where even though a galaxy property may not be predictive of the value of $f_{\rm esc}$, the detection fractions of LyC emitters may be higher when certain conditions are met (e.g. high ${\rm O_{32}}$, high EW(H$\beta$), etc.). JWST data will be key for determining whether such conditions are more often met in the high-redshift Universe compared to locally \citep[e.g.][]{Katz_2023,Cameron2023}.

\section{Discussion}
\label{sec:discussion}
In this section, we contextualize numerous indirect observational diagnostics for $f_{\rm esc}$ and develop a new model that can be used to quantitatively infer $f_{\rm esc}$ from high-redshift observations.

\subsection{Contextualising Existing Diagnostics}
\label{sec:context}
Using the theoretically motivated model for LyC leakage presented in Section~\ref{sec:framework} we proceed to predict whether individual indirect $f_{\rm esc}$ diagnostics suggested in the literature work well based on whether they match our three criteria. 

\subsubsection{$\mathrm{O}_{32}$}

There remains significant debate in the literature on the applicability of ${\rm O_{32}}$ as a diagnostic of $f_{\rm esc}$. While 1D ISM models \citep[e.g.][]{Jaskot_2013,Nakajima_2014} and certain observations \citep[e.g.][]{Faisst2016,Izotov_2018} indicate a strong correlation between O$_{32}$ and $f_{\rm esc}$, numerous theoretical models \citep[e.g.][]{Katz_2020,Barrow2020} and other observations \citep{Bassett2019,Nakajima2020,Flury_2022b} demonstrate that there are complications with viewing angle, ionization parameter, and metallicity such that high ${\rm O_{32}}$ is perhaps a necessary but insufficient condition for LyC leakage. Within our framework, ${\rm O_{32}}$ is particularly complex.

\begin{figure*}
	\includegraphics{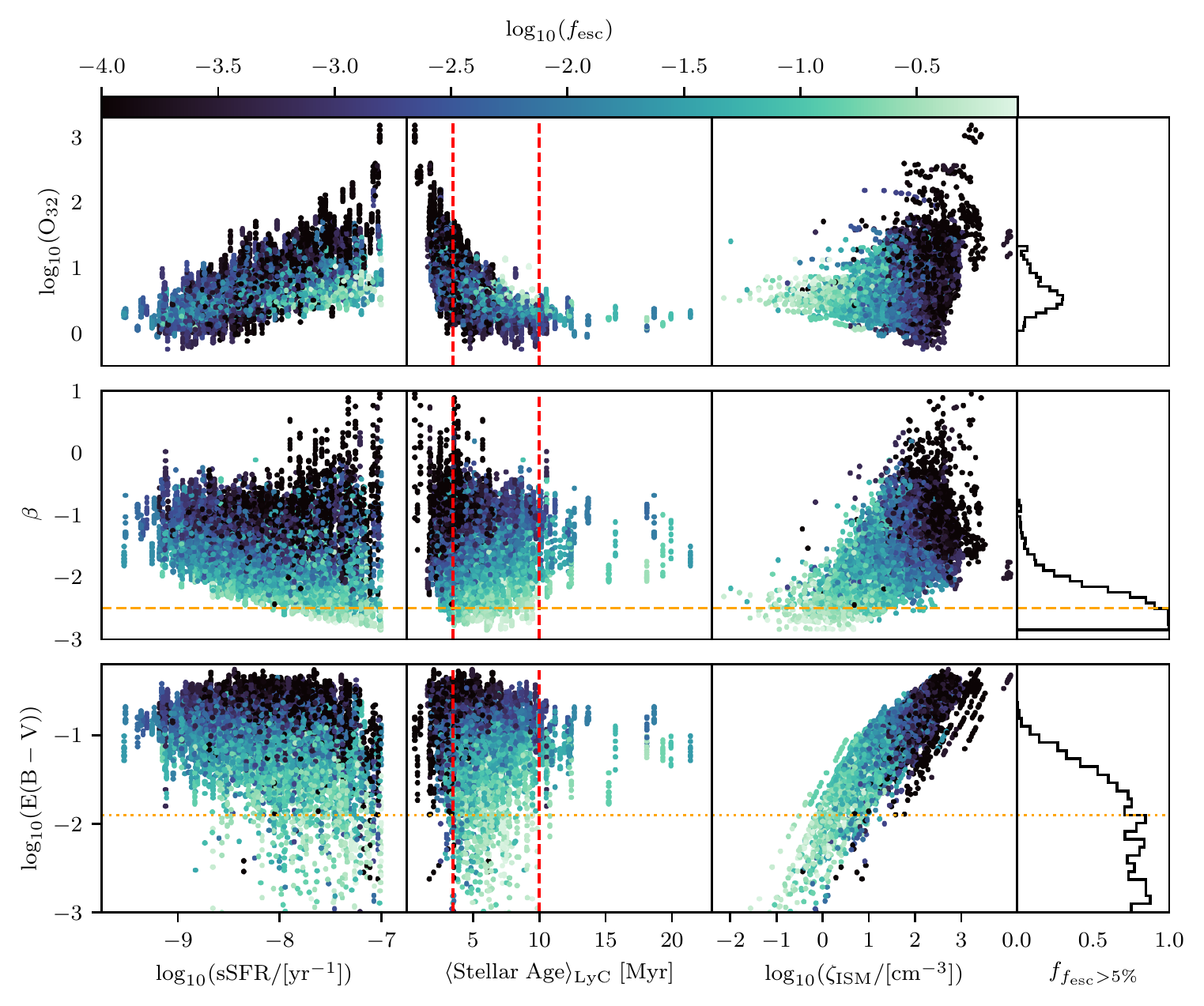}
   \caption{(Top) Observed dust-corrected (by the Balmer decrement) O$_{32}$ as a function of sSFR (left), LyC luminosity-weighted mean stellar age (second) and neutral gas attenuation parameter $\zeta_{\rm ISM}$ (third) coloured by the LyC escape fraction. The histogram (right) shows the fraction of galaxies for a given ${\rm O_{32}}$ with $f_{\rm esc} > 5\%$ for each bin which contains at least 5 such galaxies. There is a preference for leakers to have $3 < \log_{10}(\mathrm{O}_{32}) < 10$. Higher values of O$_{32}$ correlate with high sSFR, yet select for younger stellar populations and do not trace the neutral gas content of the ISM. Hence, O$_{32}$ by itself does not reliably predict $f_{\rm esc}$. (Middle) Same as above, but for spectral index $\beta$. Values of $\beta < -2.5$ are highlighted (dashed orange line) as they fulfill all three criteria, empirically selecting for high sSFR, ideal stellar age, and lower neutral gas densities in the ISM. These galaxies are the strongest leakers, as can be seen in the histogram. (Bottom) Same as above, but for the UV attenuation ${\rm E(B-V)}$. Values of ${\rm E(B-V)}$$~<0.01$ are highlighted (dotted orange line) as they appear to fulfill 2/3 criteria, empirically selecting for galaxies with the correct mean stellar population age and low $\zeta_{\rm ISM}$.}
    \label{fig:big_three_criteria_1}
\end{figure*}

As an ionization parameter diagnostic \citep[e.g.][]{Kewley2002}, high ${\rm O_{32}}$ is likely indicative of high sSFRs as seen in the top-left panel of Figure~\ref{fig:big_three_criteria_1}. There is significant scatter in the relation based on geometric effects, as well as variations with metallicity and other ISM properties, nevertheless, the highest observed values of ${\rm O_{32}}$ in our simulation traces the highest sSFR, and hence ${\rm O_{32}}$ satisfies our first criteria.

${\rm O_{32}}$ is expected to strongly vary with stellar cluster age due to the evolution of the ionizing sources (both in terms of brightness and spectral shape). This behaviour is shown in Figure~2 of \cite{Barrow2020} and is sensitive to the chosen SED model as well as the presence of Wolf-Rayet stars. The galaxies with the highest ${\rm O_{32}}$ also have the youngest LyC luminosity-weighted stellar ages (see the top-second panel of Figure~\ref{fig:big_three_criteria_1}). ${\rm O_{32}}$ peaks at ages $<3.5$~Myr and thus fails our second criterion as it preferentially traces objects that have yet to surpass the SN time scale. Similarly, ${\rm O_{32}}$ is not an indicator of neutral ISM density and only marginally traces dust, thus also failing our third criteria (see the top-third panel of Figure~\ref{fig:big_three_criteria_1}). For these reasons, by itself, ${\rm O_{32}}$ is not a good predictor of $f_{\rm esc}$.

\begin{figure}
	\includegraphics{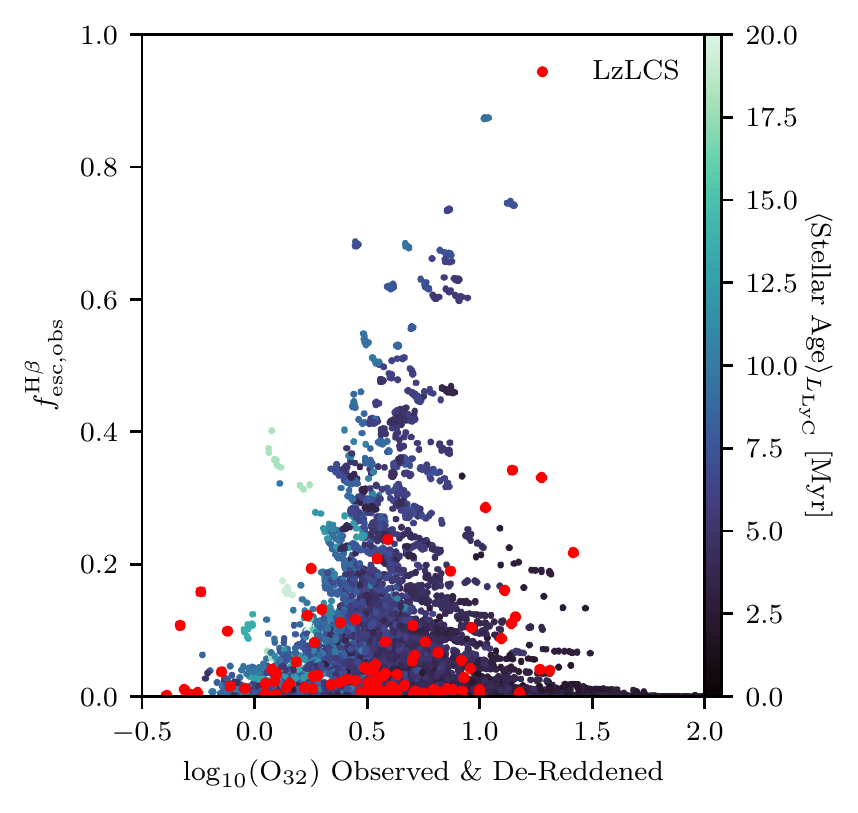}
    \caption{LyC escape fraction as a function of observed and de-reddened O$_{32}$, coloured by the ionizing luminosity weighted mean stellar age in each galaxy. Over-plotted are observational results from the LzLCS. We find that LyC escape fractions peak qualitatively at values of $\log_{10}({\rm O_{32}})$ between 3 and 10. Larger values of O$_{32}$ are dominated by young stellar populations which have yet to disrupt the ISM, producing low escape fractions.}
    \label{fig:O32_fesc_obs}
\end{figure}

In Figure~\ref{fig:O32_fesc_obs} we show the observed and de-reddened ${\rm O_{32}}$ versus $f_{\rm esc,obs}^{\mathrm{H\beta}}$. As expected, no obvious trend emerges. We find reasonably good agreement with LzLCS (red points) in that there is a preference for galaxies with high $f_{\rm esc}$ to be biased towards high ${\rm O_{32}}$, but high ${\rm O_{32}}$ does not imply high $f_{\rm esc,obs}^{\mathrm{H}\beta}$. Hence this confirms our previous assertion that high ${\rm O_{32}}$ is a necessary but insufficient condition for high $f_{\rm esc}$. We argue that the reason why there is a bias for leakers to have high observed ${\rm O_{32}}$ is two-fold. First there is a clear, albeit with significant scatter, correlation between ${\rm O_{32}}$ and sSFR which represents one of our three conditions. Second, for galaxies older than 3.5~Myr, high ionization parameter can help make feedback more efficient. Nevertheless, there are a few galaxies with high $f_{\rm esc}$ and lower ${\rm O_{32}}$ compared to the typical SPHINX$^{20}$ galaxy. The majority of these galaxies appear as light blue points on Figure~\ref{fig:O32_fesc_obs} because they have the oldest stellar ages. Such galaxies tend to be intrinsically much fainter than the galaxies with high $f_{\rm esc}$ and high ${\rm O_{32}}$ and represent the population of post-starburst Remnant Leakers described in \cite{Katz:2023b}.

We highlight that there seems to be a preferred ${\rm O_{32}}$ between 3 and 10 where galaxies are most likely to have significant LyC leakage. This range corresponds to the typical values of ${\rm O_{32}}$ that a galaxy reaches when it has evolved past the SN timescale, and can be easily seen in the top histogram of Figure \ref{fig:big_three_criteria_1}. The exact details of this peak are sensitive to Wolf-Rayet modelling as well as dust treatment, ISM gas densities, and star formation model in the simulation; all of which will need to improve before we are confident in the robustness of this particular scale. 

\subsubsection{Spectral Index $\beta$}

Recently, \cite{Chisholm_2022} have suggested that the UV slope, $\beta$, is a strong indicator of $f_{\rm esc}$, such that galaxies with bluer $\beta$ have higher $f_{\rm esc}$. $\beta$ is a readily observable quantity in the high-redshift Universe as it can be estimated for large samples of galaxies from both photometry or more accurately with JWST spectroscopy \citep[e.g.][]{Topping_2022,Endsley2022,Cullen2023}. Hence, should it be a good diagnostic of $f_{\rm esc}$, $\beta$ may be an exciting probe of LyC leakeage directly in the EoR.

While $\beta$ is not necessarily a strong indicator of sSFR, in order to have a steep slope, the young stellar population must outshine the older stellar populations in the galaxy. We empirically find (middle-left panel of Figure~\ref{fig:big_three_criteria_1}) that galaxies with the bluest $\beta$ (i.e. $<-2.5$) are also very strongly biased towards having the highest sSFRs. A high sSFR does not guarantee a blue $\beta$ and there exists strong scatter due to dust; however as an $f_{\rm esc}$ indicator, $\beta$ satisfies our first criteria. 

The UV slope is predicted to stay approximately constant for the first 10~Myr of evolution \citep{Stanway2016}. Hence a blue $\beta$ does not reveal much about the stellar population age apart from the fact that it may still be producing significant quantities of ionizing photons. Thus $\beta$ marginally satisfies our second criteria in that it picks out galaxies that can contribute to reionization but we expect some contamination from low $f_{\rm esc}$ galaxies with stellar ages younger than 3.5~Myr (middle-second panel of Figure~\ref{fig:big_three_criteria_1}). 

Finally, the observed $\beta$ is strongly sensitive to dust. While not a density indicator, $\beta$ can easily select for galaxies with very low ${\rm E(B-V)}$ and hence $\beta$ also marginally satisfies our third criteria. For these reasons, in a metal-enriched environment we expect $\beta$ to be a relatively good indicator of $f_{\rm esc}$, albeit with significant scatter. 

\begin{figure}
	\includegraphics{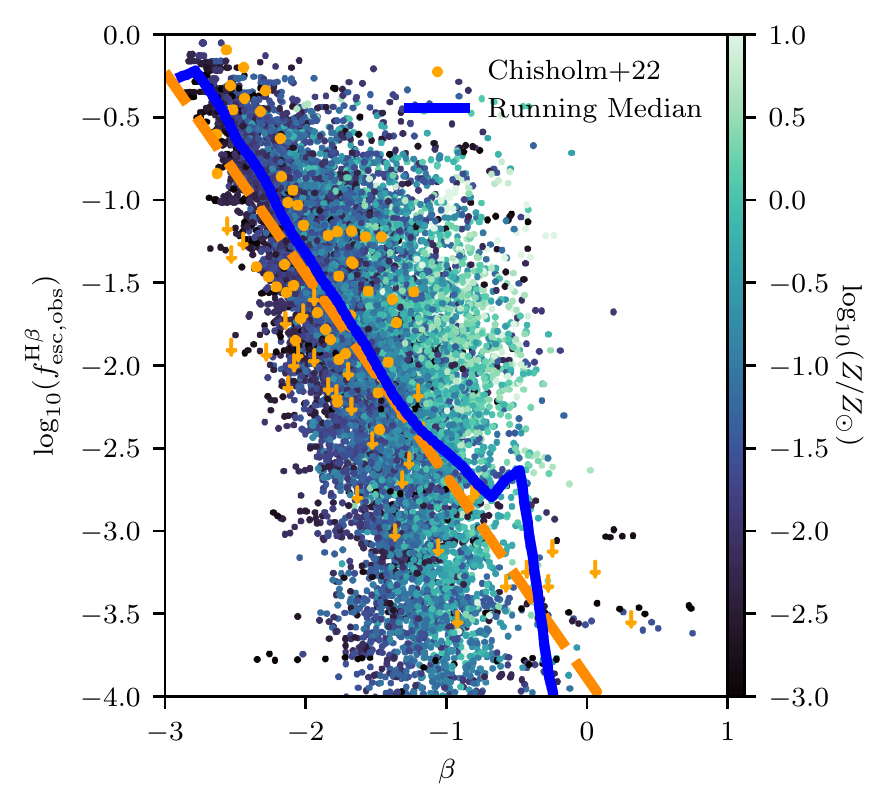}
   \caption{LyC escape fraction as a function of observed spectral index $\beta$ and coloured by gas metallicity. Over-plotted are observational data-points and predicted relation (orange) from \protect\cite{Chisholm_2022}, as well as our median relation (blue). We find that our data agrees well with observational data, confirming that $\beta$ strongly anti-correlates with LyC escape fraction.}
    \label{fig:fesc_beta}
\end{figure}

In Figure~\ref{fig:fesc_beta} we plot $f_{\rm esc,obs}^{\mathrm{H}\beta}$ versus the observed $\beta$ for SPHINX$^{20}$ galaxies and we find a strong trend that high $f_{\rm esc,obs}^{\mathrm{H}\beta}$ galaxies tend to have bluer $\beta$ (as can be clearly seen in the middle histogram of Figure \ref{fig:big_three_criteria_1}). We do find systems that also have redder $\beta$ and high $f_{\rm esc,obs}^{\mathrm{H}\beta}$ but some of this is a sight-line effect and likewise, Remnant Leakers, which do not contribute meaningfully to reionization, will populate this region of the plot. The orange points and dashed orange line show both the data and relation predicted by \cite{Chisholm_2022} which is in very good agreement with our median relation (blue).

We emphasize that this result holds only when metals/dust are present as without dust obscuration, our third criterion would not be satisfied. Indeed the galaxies with the lowest $f_{\rm esc}$ at the bluest $\beta$ are the most metal-poor galaxies in our simulated sample. Our model naturally predicts that this could become problematic at $Z/Z_{\odot}<0.01$. Here we have assumed that the dust-to-gas mass ratio scales linearly with metallicity; however, observations show that dust content may fall off as a power-law with decreasing metallicity \citep[e.g.][]{Remy-Ruyer_2014}. In that case, we expect that the critical metallicity where $\beta$ no longer becomes a good $f_{\rm esc}$ indicator occurs at a higher metallicity than $Z/Z_{\odot} = 0.01$. Hence to be conservative, we advocate that observed $\beta$ is likely to be a good $f_{\rm esc}$ indicator at $Z/Z_{\odot}>0.1$ and the details at lower metallicity (and the ability to apply this relation in the epoch of reionization) are sensitive to how the dust-to-gas mass ratio scales with metallicity and the timescales of dust formation at high-redshift.

\subsubsection{${\rm E(B-V)}$}

The problem of whether dust attenuation strongly affects $f_{\rm esc}$ is not completely understood. \cite{Chisholm_2018} suggest that even small dust attenuation removes significant numbers of ionizing photons. However, simulations have shown that dust tracks neutral hydrogen which has a much more important impact on $f_{\rm esc}$ \citep{Katz_2022b}. Therefore, it is natural to explore the use of UV attenuation (e.g. ${\rm E(B-V)}$, derived from the Balmer decrement) as a potential $f_{\rm esc}$ diagnostic. \cite{Saldana-Lopez2022} found a strong anti-correlation between the UV dust-attenuation and LyC escape fraction for galaxies in the LzLCS sample, suggesting that LyC leakers tend to have a dust-poor ISM.

Unsurprisingly, we find no significant dependence of ${\rm E(B-V)}$ on sSFR (bottom-left panel of Figure~\ref{fig:big_three_criteria_1}), with any residual relationship introduced by the fact that stellar mass is the denominator of sSFR and high-mass galaxies tend to be more metal-enriched and thus have more dust. ${\rm E(B-V)}$ does not pass our first criterion. Interestingly, we find that galaxies outside our stellar age criterion tend to have significantly more UV attenuation (bottom-second panel of Figure~\ref{fig:big_three_criteria_1}). This points to the fact that SNe are able to destroy dust (in our case by destroying neutral hydrogen) through mechanical feedback \citep[e.g.][]{Priestley_2021}. Therefore, ${\rm E(B-V)}$ marginally satisfies our second criterion. Finally, it is clear that (by construction), there is a strong correlation between ${\rm E(B-V)}$ and $\zeta_{\rm ISM}$ (bottom-third panel of Figure~\ref{fig:big_three_criteria_1}). Therefore, we conclude that ${\rm E(B-V)}$ should be a good indicator of LyC leakage.

\begin{figure}
	\includegraphics{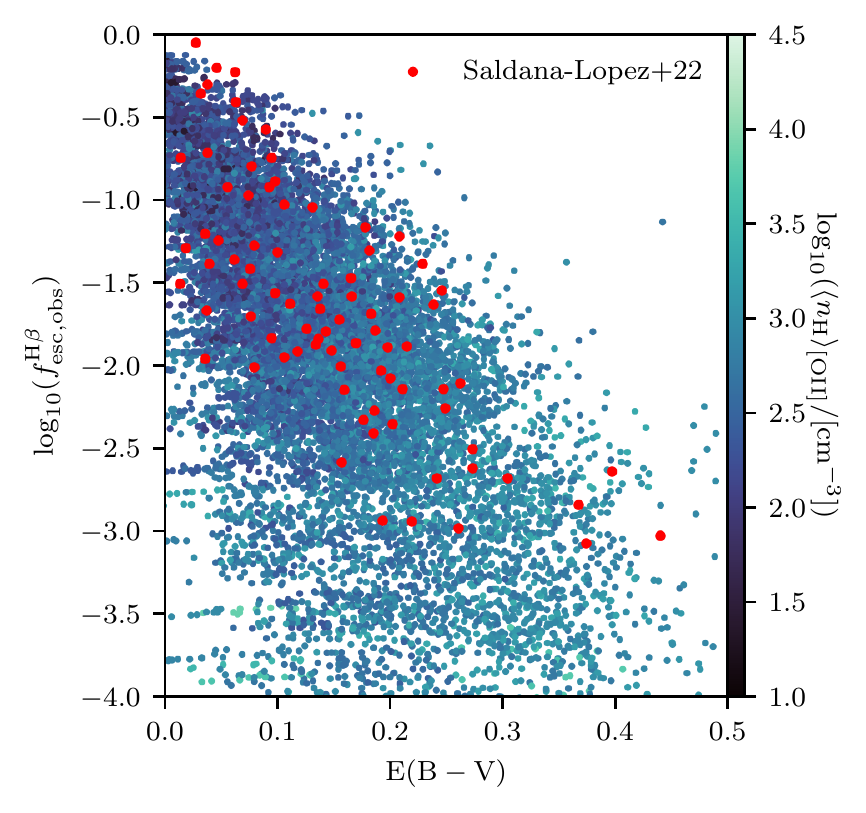}
   \caption{Observed LyC escape fraction as a function of dust-attenuation at 912~\angstrom~coloured by the mean neutral gas density weighted by the O~{\small II} doublet ratio, compared to data from the LzLCS \citep{Saldana-Lopez2022}.}
    \label{fig:fesc_ebmv}
\end{figure}

In Figure~\ref{fig:fesc_ebmv} we show the observed $f_{\rm esc}$ as a function of ${\rm E(B-V)}$ compared to LzLCS galaxies from \cite{Saldana-Lopez2022}. We find a strong trend between the two quantities. Furthermore, galaxies with lower $\langle n_{\rm H}\rangle_{[\mathrm{OII}]}$ exhibit less scatter. However, it is likely that such comparisons are sensitive to the dust model used in the simulation. Similarly, \cite{Saldana-Lopez2022} assume a uniform dust screen which is not representative of the dust distribution in our simulation. 

\subsubsection{$\Sigma_{\rm SFR}$}

Star formation rate surface density is perhaps intuitively a good indicator of $f_{\rm esc}$. Since $f_{\rm esc}$ is predicted to be feedback-regulated \citep[e.g.][]{Trebitsch_2017,Kimm_2017}, concentrated star formation may help increase the local efficiency of mechanical feedback, creating optically thin, low-density channels in the ISM. With limited empirical data, \cite{Sharma2017,Naidu_2020} assumed that galaxies with the highest $\Sigma_{\rm SFR}$ also have the highest $f_{\rm esc}$, which can result in a reionization scenario dominated by the most massive galaxies. 

\begin{figure*}
	\includegraphics{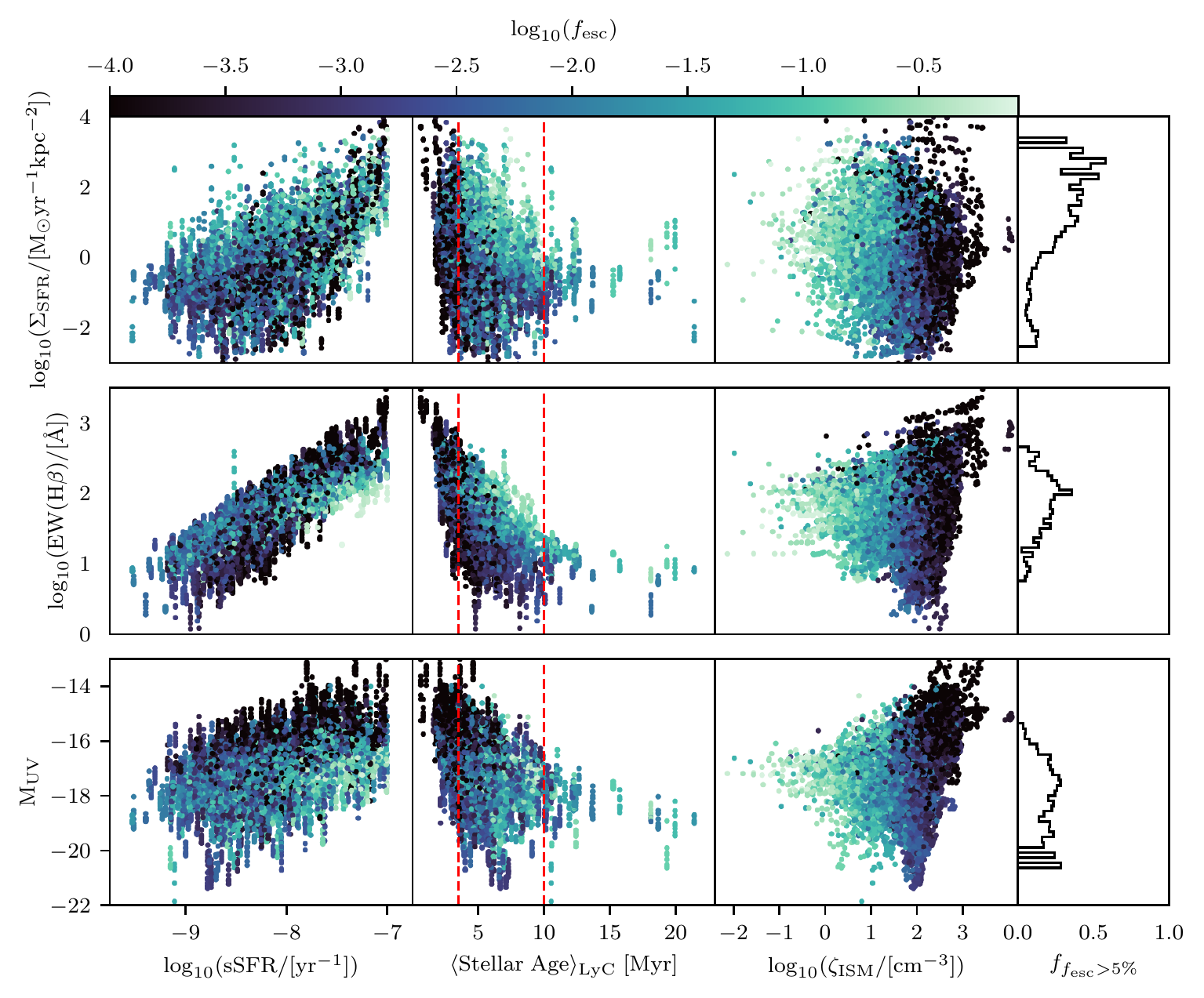}
   \caption{(Top) Observed $\Sigma_{\rm SFR}$ (as defined in Section \ref{sec:methods}) as a function of sSFR (left), LyC luminosity-weighted mean stellar age (second) and neutral gas attenuation parameter $\zeta_{\rm ISM}$ (third) coloured by the LyC escape fraction. The histogram (right) shows the fraction of galaxies for a given $\Sigma_{\rm SFR}$ with $f_{\rm esc} > 5\%$ for each bin that contains at least 5 such galaxies. Though $\Sigma_{\rm SFR}$ correlates weakly with sSFR (as both depend explicitly on SFR), it selects for young stellar ages and does not trace $\zeta_{\rm ISM}$. Therefore, $\Sigma_{\rm SFR}$ by itself is not a reliable diagnostic for the LyC escape fraction.
   (Middle) Same as above, but for H$\beta$ equivalent widths. EW(H$\beta$) traces sSFR very well, but greater values select for stellar ages $< \SI{3.5}{\mega\year}$ and do not trace the state of the ISM. We therefore expect no strong relation with the LyC escape fraction, though values of 100~\angstrom~are weakly preferred. 
   (Bottom) Same as above, but for M$_{\rm UV}$. We find that M$_{\rm UV}$ does not trace the sSFR, but bright magnitudes select for the correct stellar ages. However, M$_{\rm UV}$ shows no dependence on ISM density. As a result, leakers show a weak bias towards brighter magnitudes, but M$_{\rm UV}$ is not a useful indicator of the LyC escape fraction.}
    \label{fig:big_three_criteria_2}
\end{figure*}

As SFR is the numerator of $\Sigma_{\rm SFR}$, there is unsurprisingly a strong correlation between sSFR and $\Sigma_{\rm SFR}$ (top-left panel of Figure~\ref{fig:big_three_criteria_2}). This is due to the weaker dependence of galaxy size on stellar mass \citep{Kawamata2018,Bouwens2022b}. As with previous diagnostics, we similarly find significant scatter due to the variable impact of dust on the H$\beta$ emission and projected galaxy size as a function of sight line. $\Sigma_{\rm SFR}$ undoubtedly passes our first criterion.

In contrast we find that the highest values of $\Sigma_{\rm SFR}$ tend to correspond to galaxies with ages younger than the SN timescale (top-second panel of Figure~\ref{fig:big_three_criteria_2}), although there are a significant number of galaxies that have ages $>3.5$~Myr and $\Sigma_{\rm SFR}>10\ {\rm M_{\odot}/yr/kpc^2}$ (the canonical value reported in \cite{Flury_2022b} as the threshold for strong leakage). Galaxy size in our simulations is also relatively stable beyond 10~Myr. Thus, $\Sigma_{\rm SFR}$ alone does not satisfy our second criterion and may be biased towards galaxies with too young stellar ages. Finally, we find no relation between $\Sigma_{\rm SFR}$ and the state of the ISM (top-third panel of Figure~\ref{fig:big_three_criteria_2}). We then might expect that galaxies with high $f_{\rm esc}$ might also have high $\Sigma_{\rm SFR}$ due to the correlation with sSFR, but we expect there to be no strong correlation. 

\begin{figure}
	\includegraphics{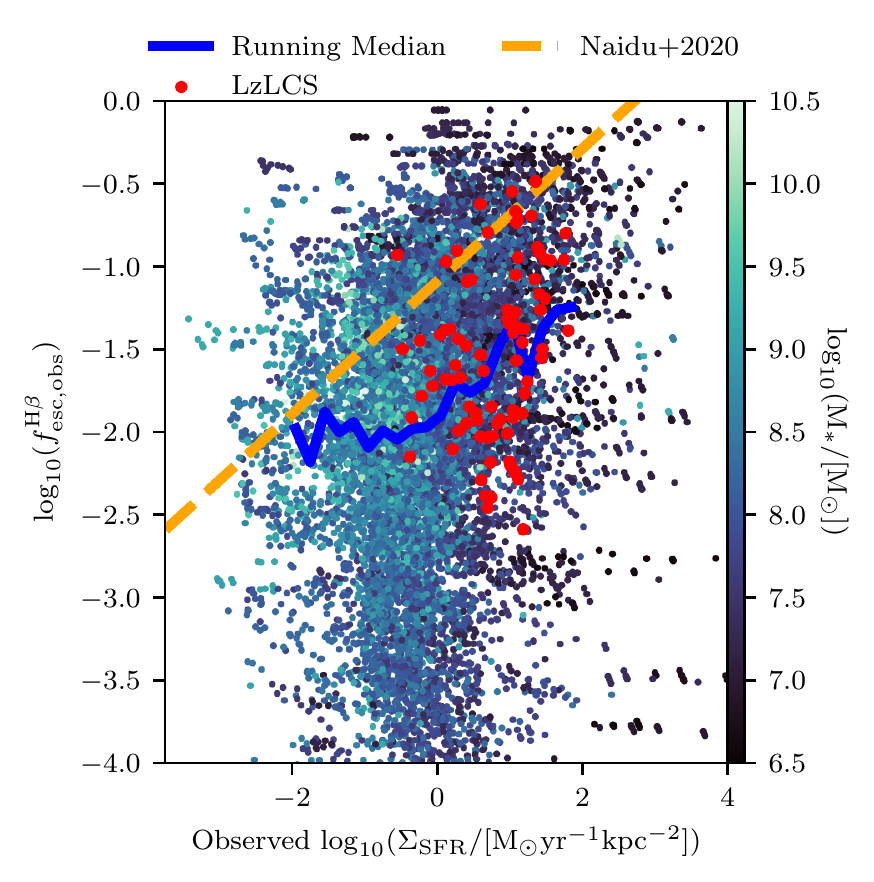}
    \caption{LyC escape fraction as a function of $\Sigma_{\rm SFR}$ coloured by stellar mass. LzLCS data is shown in red, while the relation suggested by \protect\cite{Naidu_2020} is shown in orange. We find that SPHINX$^{20}$ galaxies agree well with LzLCS observations, while disagreeing with the proposed relation of \protect\cite{Naidu_2020}.}
    \label{fig:fesc_sigmasfr}
\end{figure}

Indeed in Figure~\ref{fig:fesc_sigmasfr} we find no trend between the quantities. The relation suggested by \cite{Naidu_2020} does not envelope our data, nor does it represent the LzLCS galaxies, which are consistent with those presented here. Similarly, the assumption by \cite{Sharma2017} that all galaxies with $\Sigma_{\rm SFR}>0.1\ {\rm M_{\odot}/yr/kpc^2}$ have $f_{\rm esc}=20\%$ is clearly an inaccurate representation of both SPHINX$^{20}$ galaxies and LzLCS. Interestingly, in our SFR-limited sample, the lowest halo mass galaxies have the highest $\Sigma_{\rm SFR}>0.1\ {\rm M_{\odot}/yr/kpc^2}$ as they tend to fall above the main sequence of star formation.

\cite{Flury_2022b} do note a trend between $\Sigma_{\rm SFR}$ and $f_{\rm esc}$. It is possible that one emerges due to their selection criteria which are not fully representative of the general galaxy population. Hence trying to reproduce their correlation coefficient with SPHINX$^{20}$ may result in the correct value for the wrong reasons. Finally, we find that by selecting with $\Sigma_{\rm sSFR}$ one is more likely to find a galaxy with significant leakage, which can be seen by comparing the histogram in Figures \ref{fig:mega_three_criteria} and the top histogram of \ref{fig:big_three_criteria_2}. This is in contrast with findings in the LzLCS, which report that the addition of stellar mass did not improve selection power \citep{Flury_2022b}.

\subsubsection{${\rm H\beta\ Equivalent\ Width}$}
\label{sec:EWHB}

The connection between H$\beta$ equivalent width (EW(H$\beta$)) and $f_{\rm esc}$ remains debated. Green Pea galaxies are among the most studied low-redshift galaxy populations that contain a significant number of LyC leakers \citep[e.g.][]{Izotov_2016,Izotov_2018,Izotov_2018b} and these systems often exhibit extreme emission line ratios and equivalent widths \citep[e.g.][]{Yang2017}. In contrast, as the escape fraction approaches $100\%$, EW(H$\beta$) should tend towards zero as none of the ionizing photons are absorbed \citep{Zackrisson_2017}. While LzLCS find no strong trend between EW(H$\beta$) and $f_{\rm esc}$, galaxies with high $f_{\rm esc}$ tend to also have high EW(H$\beta$). 

EW(H$\beta$) is a very strong tracer of sSFR (middle-left panel of Figure~\ref{fig:big_three_criteria_2}) as Balmer lines have long been known to track SFR and the strength of the continuum is sensitive to total stellar mass \citep[e.g.][]{Kennicutt1998}. However, like ${\rm O_{32}}$, the highest values of EW(H$\beta$) trace ages $<3.5$~Myr and there is no correlation between EW(H$\beta$) and the state of the ISM (middle-second and middle-third panels of Figure~\ref{fig:big_three_criteria_2}) Hence we expect no strong correlation between EW(H$\beta$) and $f_{\rm esc}$; although, the connection with sSFR would explain the trend seen in LzLCS that the LyC leaker fractions increases with EW(H$\beta$). We note that similar to our findings for O$_{32}$, there appears to be a characteristic EW(H$\beta$) $\sim 100$\angstrom~ for which galaxies tend to show elevated $f_{\rm esc}$ (see the middle histogram of Figure \ref{fig:big_three_criteria_2}). However, for the same reasons as before it is difficult to claim a robust value.

\subsubsection{$\mathrm{M}_{\mathrm{UV}}$}
\label{sec:MUV}

\cite{Flury_2022b} recently reported a weak correlation between $f_{\rm esc}$ and M$_{\rm UV}$, in agreement with other observations that indicate LyC leakers tend to be lower mass, fainter galaxies \citep{Steidel_2018,Pahl_2021}. In contrast, \citep{Rosdahl_2022,Saxena_2022} find no relation between between $f_{\rm esc}$ and M$_{\rm UV}$, although they agree that lower luminosity galaxies are likely the sources of reionization. 

\begin{figure}
	\includegraphics{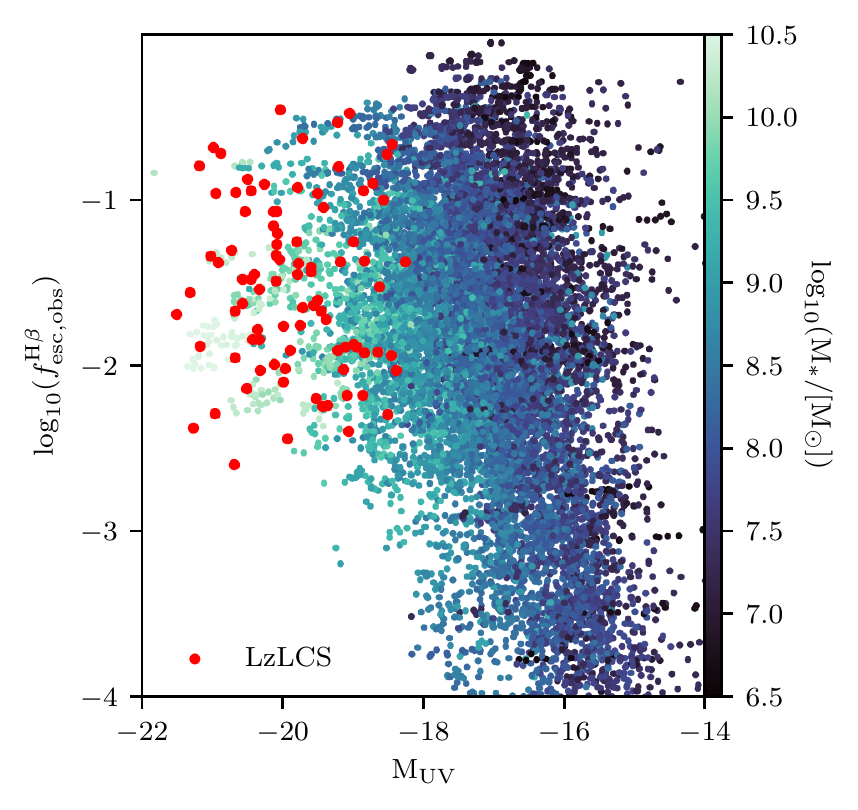}
   \caption{LyC escape fraction as a function of line-of-sight M$_{\rm UV}$ coloured by the stellar mass. Data from LzLCS is shown in red, agreeing with bright SPHINX$^{20}$ galaxies. We find that when all masses are considered, M$_{\rm UV}$ is not a reliable diagnostic for the LyC escape fraction.}
    \label{fig:fesc_MUV}
\end{figure}

In general, M$_{\rm UV}$ is a weak indicator of sSFR (bottom-left panel of Figure \ref{fig:big_three_criteria_2}) while high M$_{\rm UV}$ could also indicate ages $<3.5$~Myr (bottom-second panel of Figure \ref{fig:big_three_criteria_2}). Similarly we find no correlation between M$_{\rm UV}$ and ISM state (bottom-third panel of Figure \ref{fig:big_three_criteria_2}) so our model would predict no correlation between M$_{\rm UV}$ and $f_{\rm esc}$ as shown in \cite{Rosdahl_2022}. Figure \ref{fig:fesc_MUV} demonstrates this, comparing our galaxies to those of the LzLCS \citep{Flury_2022b}. Here, we find that the majority of this scatter is introduced by a strong dependence of M$_{\rm UV}$ on stellar mass.

\begin{figure}
	\includegraphics{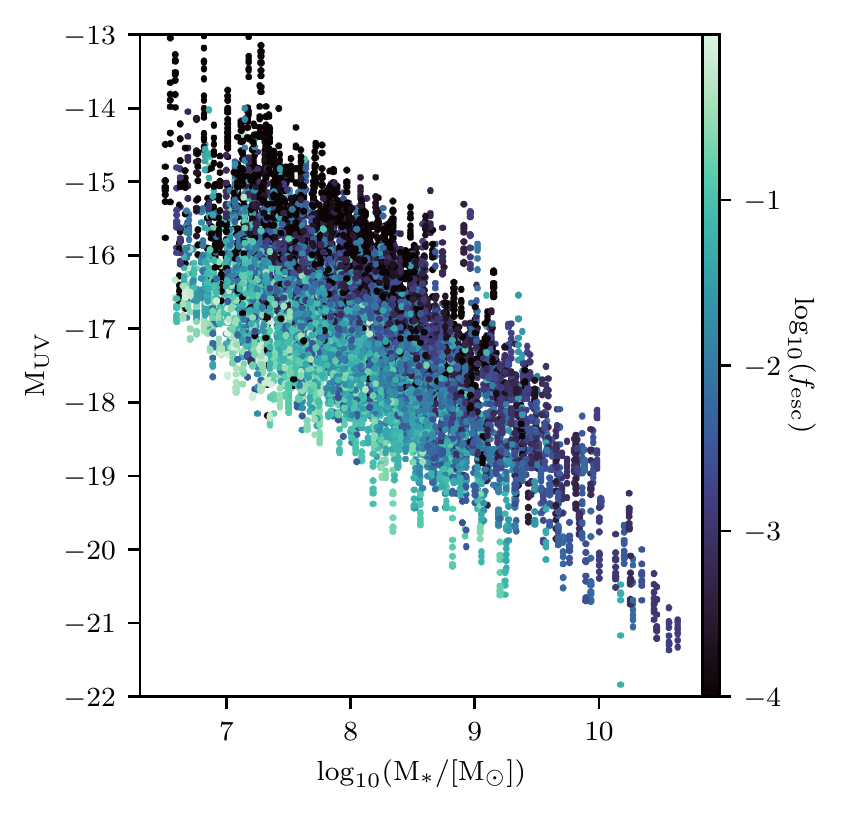}
   \caption{Line-of-sight observed UV magnitude for each mock observation as a function of the galaxy stellar mass, coloured by the LyC escape fraction. We find that more massive galaxies tend to be more luminous in the UV, and that for a fixed galaxy mass bin, brighter galaxies tend to have higher LyC escape fractions. Therefore, at constant stellar mass, M$_{\rm UV}$ can be used as a LyC diagnostic.}
    \label{fig:mass_MUV}
\end{figure}

Figure~\ref{fig:mass_MUV} explores this further, by comparing M$_{\rm UV}$ with M$_{\rm *}$. Points are coloured by $f_{\rm esc}$ such that brighter galaxies at fixed stellar mass are biased towards having higher $f_{\rm esc}$. When stellar mass is fixed, high intrinsic M$_{\rm UV}$ correlates with a high sSFR. However, like the observed $\beta$ at fixed stellar mass, observed M$_{\rm UV}$ becomes a strong indicator of dust attenuation which represents one of the tracers of our combination parameter on ISM state. Thus at fixed stellar mass the observed M$_{\rm UV}$ satisfies two out of three criteria with scatter introduced due to stellar age. This demonstrates that sample selection is key for the emergence of trends between $f_{\rm esc}$ and various galaxy properties.

\subsection{A New Combined Diagnostic}

Inspired by the success of our theoretical selection criteria in Figure \ref{fig:three_crit_sel} (top), we now aim to reproduce the ability of a three-point criterion at isolating systems with high LyC leakage, albeit with quantities which are directly observable. Furthermore, we can make use of the fact that several quantities satisfy multiple criteria to find the best set of observables from which to construct our diagnostic.

Given the fact that galaxies with blue dust-attenuated $\beta$ marginally satisfy criteria 1 and 2, while strongly satisfying 3 (see middle row of Figure \ref{fig:big_three_criteria_1}) we choose to start here. We continue with our second-strongest individual diagnostic, ${\rm E(B-V)}$, which weakly indicates criterion 2 and strongly satisfies criterion 3 (see bottom row of Figure \ref{fig:big_three_criteria_1}). Finally, within this set of diagnostics we note that the most loosely constrained is mean stellar age, for which we now select the H$\alpha$-to-FUV flux ratio, given the fact that it has been previously shown to indicate stellar age both observationally \citep{Weisz_2012} and in simulations \citep{Sparre_2017}, though this is by no means settled \cite[c.f.][]{Rezaee_2022}. Therefore, we predict that this set of three diagnostics can be used to reliably select a sample of galaxies which are greatly enriched with LyC leakers.

\begin{figure}
    \includegraphics{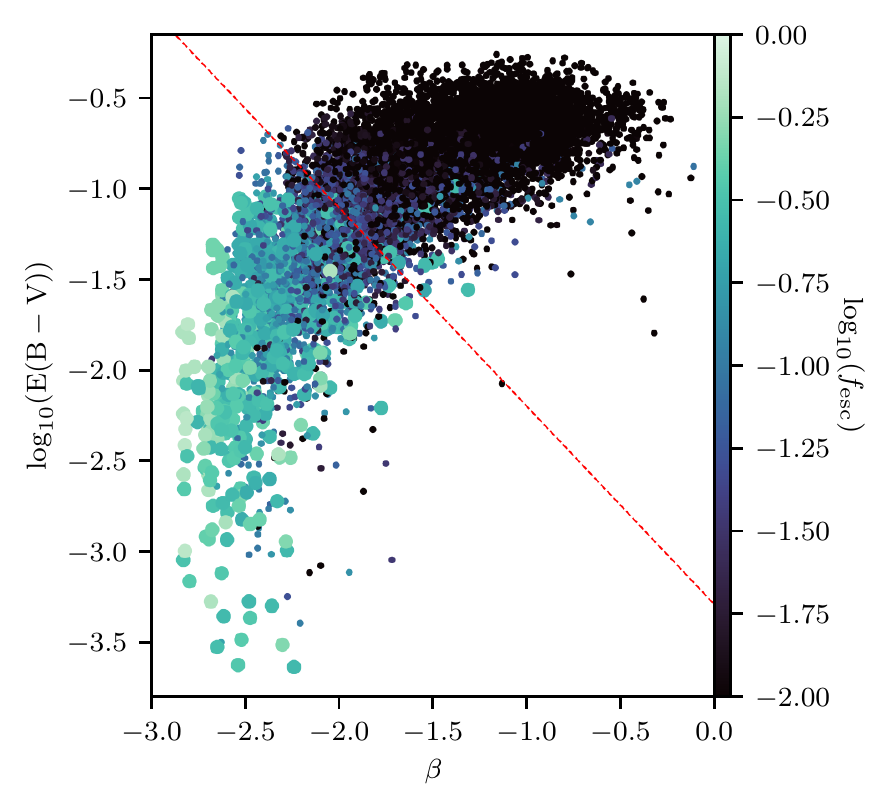}
    \caption{Line-of-sight measurements of $\beta$ as a function of ${\rm E(B-V)}$ for all observations of galaxies in our sample with observed $\log_{10}(\mathrm{H}\alpha/\mathrm{L}_{1500}) < 1.6$ coloured by the true LyC escape fraction. Systems with escape fractions greater than 20\% are enlarged. A selection has been made (red, given by Equation \ref{eq:observational_cut}) to produce a sample of highly enriched leakers, 67\% of which have $f_{\rm esc} > 5\%$. Those selected by this set of criteria account for 62\% of the total population of such systems in our sample.}
    \label{fig:three_crit_sel_obs}
\end{figure}

Figure \ref{fig:three_crit_sel_obs} shows this, plotting line-of-sight $\beta$ as a function of ${\rm E(B-V)}$ for all galaxies with $\log_{10}(\mathrm{H}\alpha/\mathrm{L}_{1500}) < 1.6$\footnote{We note that while using a stronger constraint, e.g. $\log_{10}(\mathrm{H}\alpha/\mathrm{L}_{1500}) < 1.4$ will produce a more enriched sample of galaxies, it will inevitably represent a smaller set of all LyC leaking systems.} coloured by $f_{\rm esc}$, with galaxies with $f_{\rm esc} > 20\%$ shown as larger points. Here, we find that by selecting galaxies with 

\begin{equation}
    \log_{10}({\rm E(B-V)}) < -1.1\beta - 3.3, \label{eq:observational_cut}
\end{equation}
we generate a sample that is highly enriched with LyC leakers. Namely, this reduced set consists of 1931 observations (16.9\% of our sample), $\sim 67\%$ of which have $f_{\rm esc} > 5\%$. This accounts for 62\% of all such galaxies in our sample. In contrast, the theoretical selection criteria given in Figure \ref{fig:three_crit_sel} is 74\% enriched by such galaxies, accounting for 65\% of the overall sample. We stress that this is a theoretically motivated set of directly observable diagnostics which successfully selects for the majority of LyC leakers in our sample.

\subsection{Predicting $f_{\rm esc}$ for bright Ly$\alpha$ emitters observed with JWST}
\label{sec:GLM}

There remains debate in the literature over the contribution of bright Ly$\alpha$ emitters (typically defined as having $L_{\rm Ly\alpha} > 10^{42.2}\;{\rm erg}/{\rm s}$ and ${\rm EW}({\rm Ly\alpha}) > 25$\angstrom) to reionization. For example, based on lower-redshift stacks, \cite{Naidu2022} assumed that bright Ly$\alpha$ emitters with low peak velocity separation and line-centre flux all have escape fractions of $20\%$. Furthermore, \cite{Matthee2022} showed that if half of bright Ly$\alpha$ emitters have an escape fraction of 50\%, then one can match observational constraints on the neutral fraction evolution. In contrast, based on a small stack of $z>7$ galaxies, \cite{Witten2023} inferred that bright Ly$\alpha$ emitters have $f_{\rm esc}\lesssim10\%$. 

One of the primary advancements of JWST compared to earlier observations is the ability to obtain high resolution spectra of the rest-frame UV and optical for large numbers of galaxies at $z>6$. While our previous discussion has focused on which indirect indicators are likely to identify enriched samples of LyC leakers, here we use the observable properties of galaxies to quantitatively predict the value of the escape fraction, particularly for a sample of high-redshift Ly$\alpha$ emitters. A similar exercise was performed in \cite{Maji2022}; however, the main difference here is that we focus only on observable quantities such that our models are immediately applicable to available galaxy spectra. 

\begin{figure}
    \includegraphics{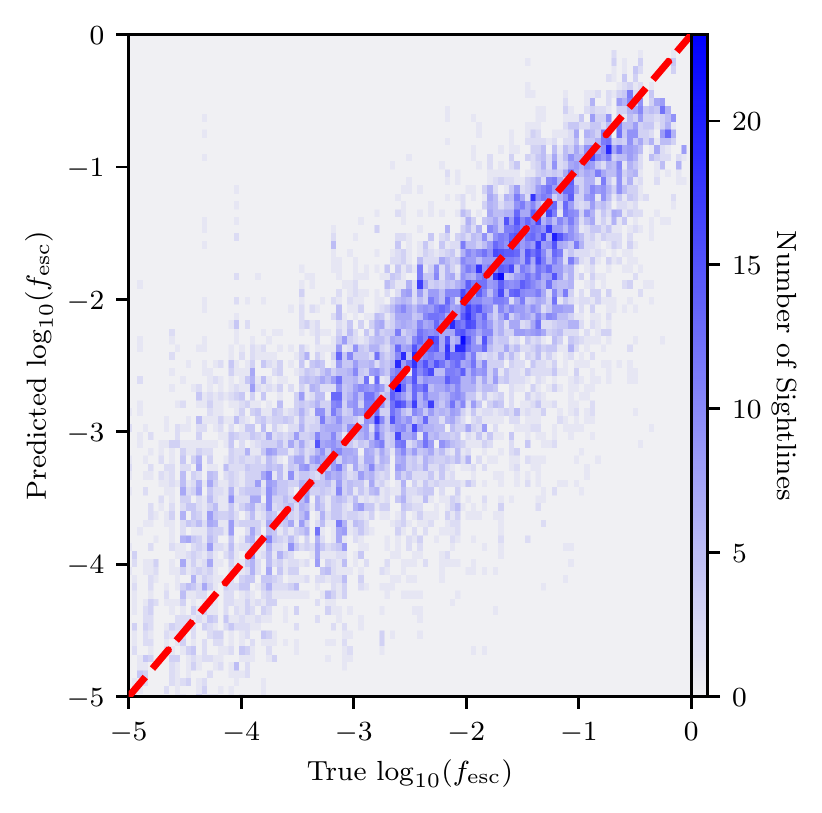}
    \caption{Histogram of predicted $f_{\rm esc}$ as a function of true $f_{\rm esc}$ for our entire sample, as estimated by the generalized linear model given by Equation \ref{eqn:fesc_predict} with parameters from Table \ref{tab:coef_fesc}. Points are coloured by the number of sightlines in each bin. The one-to-one relation is shown in red.}
    \label{fig:GLM}
\end{figure}

We consider eight observable quantities: $\beta$, ${\rm E(B-V)}$, H$\beta$, EW(H$\beta$), M$_{\rm UV}$, ${\rm R_{23}}$, ${\rm O_{32}}$, and the half-light radius measured at 1500~\angstrom. We then run a generalized linear model on scaled data\footnote{i.e. mean of zero and unit variance.} using L1 regularization to limit the number of needed measurements and maintain simplicity so that the model is interpretable. To avoid over-fitting, we have split\footnote{When we split the data, we ensure that all viewing angles for each galaxy are part of the same class to avoid information leakage.} the data such that the model is trained on 80\% and the remaining 20\% is used for validation. Six of the eight initial parameters have non-zero coefficients, with both EW(H$\beta$) and the half-light radius proving irrelevant for our model. Using these six parameters, $f_{\rm esc}$ can be estimated as:
\begin{equation}
\label{eqn:fesc_predict}
\log_{10}\left(f_{\rm esc}\right) = -2.5 + \sum_{i=1}^6 C_i \frac{p_i-\bar{p}_i}{\sigma_{p_i}}.
\end{equation}
Values for all coefficients are listed in Table~\ref{tab:coef_fesc}. The median absolute error on the training and validation sets are identical at 0.39~dex indicating that the model generalizes well. This can also be seen in Figure \ref{fig:GLM}, where we show predicted $f_{\rm esc}$ as a function of true $f_{\rm esc}$ for our sample.

\begin{table}
	\centering
	\caption{Coefficients and constants required to solve Equation~\ref{eqn:fesc_predict} to quantitatively predict the value of $f_{\rm esc}$ from an observed spectrum.}
	\label{tab:coef_fesc}
	\begin{tabular}{llrrr} 
		\hline
		$i$ & $p_i$ & $\bar{p}_i$ & $\sigma_{p_i}$ & $C_i$ \\
        \hline
        1 & $\beta$ & $-1.528$ & 0.526 & $-0.641$ \\
        2 & ${\rm E(B-V)}$ & 0.139 & 0.094 & $-0.391$ \\
        3 & $\log_{10}({\rm H\beta/[erg\ s^{-1}]})$ & 40.630 & 0.342 & $0.030$ \\
        4 & M$_{\rm UV}$ & $-17.055$ & 1.110 & $-0.021$ \\
        5 & $\log_{10}({\rm R_{23}})$ & 0.661 & 0.235 & $0.031$ \\
        6 & $\log_{10}({\rm O_{32}})$ & 0.680 & 0.357 & $-0.424$ \\
		\hline
	\end{tabular}
\end{table}

The coefficients provide insight into how each parameter correlates with $f_{\rm esc}$. For example, the coefficients for $\beta$ and ${\rm E(B-V)}$ are strongly negative indicating that blue UV slopes and low dust content are strong indicators of leakage. Counter-intuitively, ${\rm O_{32}}$ negatively scales with $f_{\rm esc}$ in our model. We believe this is due to the fact that once galaxies with blue UV slopes and low dust are selected, ${\rm O_{32}}$ becomes an age indicator and lower ${\rm O_{32}}$ indicates older ages. We emphasize that this anti-correlation exists only in the framework that contains these other parameters.

As a first application, we apply this relation to JADES-GS-z7-LA, a $z=7.3$ Ly$\alpha$ emitter \citep{Saxena2023} recently discovered as part of the JWST JADES GTO program. Based on a variety of spectroscopic features, the authors concluded that the current value of $f_{\rm esc}$ is not substantially high, despite the high Ly$\alpha$ $f_{\rm esc}$. Using our model, we predict a value of 3\% which is considerably lower than the value needed to inflate an ionized bubble such that Ly$\alpha$ is not completely attenuated by the IGM \citep{Saxena2023}. Hence it is more likely that faint nearby dwarf galaxies (such as those which have already been spectroscopically confirmed, see Figure 5 of \citealt{Witstock:2023}) are likely responsible for the local ionized bubble. Furthermore, Equation \ref{eqn:fesc_predict} has been used to infer $f_{\rm esc}$ for 16 faint Ly$\alpha$ emitters at $z > 5.8$ \citep{Saxena:2023b}.

Beyond $z\sim9.5$ [O~{\small III}]~$\lambda$5007 drops out of NIRSpec; however, GN-z11, a spectroscopically confirmed $z=10.6$ galaxy was recently discovered to be a bright Ly$\alpha$ emitter \citep{Bunker2023}. Because H$\beta$, ${\rm O_{32}}$, and ${\rm R_{23}}$ have not been measured for this galaxy, we create a custom model using H$\gamma$ and [O~{\small III}]~$\lambda$4363 rather than [O~{\small III}]~$\lambda$5007. We find an escape fraction of 11\%, significantly greater than the 4\% reported as the Ly$\alpha$ escape fraction. While it is theoretically difficult to obtain a LyC escape fraction greater than that of Ly$\alpha$ and similarly this is rarely observed \citep[e.g.][]{Izotov_2022,Verhamme_2017}, there is undoubtedly scatter in our model, which more than accounts for this discrepancy. IGM attenuation can also play a role in extinguishing the observed Ly$\alpha$. \cite{Hayes2023} argue that the  Ly$\alpha$ escape from GN-z11 into the IGM is as high as 50\%, which makes our measured LyC escape fraction fully consistent. Similarly the SED fit for GN-z11 shows a marginal $A_v=0.17$. If we input this into our model (rather than the fiducial parameters that assumed no dust as suggested in \citealt{Bunker2023}), the estimated LyC escape fraction decreases to 6\%. It is important to note that the [O~{\small III}]~$\lambda$4363 line contains slightly different information than [O~{\small III}]~$\lambda$5007 on account of it being an auroral line and therefore  may be a useful tool in searching for LyC leakers. However, we have chosen not to include it in our general model on account of the fact that it is generally difficult to detect, particularly at high redshifts \citep[e.g.][]{Laseter:2023}.

While these two galaxies represent single-object detections, \cite{Tang2023} recently published a sample of six $z>7$ Ly$\alpha$ emitters from CEERS \citep{Finkelstein2022}. Among these six, four galaxies (CEERS-1019, CEERS-1027, CEERS-698, \& CEERS-44) have measurements for all of the quantities we need to apply our model. We find LyC escape fractions of 4\%, 0.6\%, 0.7\%, and 10\%, respectively. Compared to the Ly$\alpha$ escape fractions for these galaxies (4\%, 9\%, 5\%, and 34\%), our measured LyC $f_{\rm esc}$ values are once again consistent with Ly$\alpha$ escape being higher than LyC escape.

In general, our model seems to point to bright Ly$\alpha$ emitters having LyC escape fractions $\lesssim10\%$. This result agrees with \cite{Witten2023} but contradicts \cite{Naidu2022} where it was assumed that bright Ly$\alpha$ emitters with low peak velocity separation and line-centre flux all have $f_{\rm esc}=20\%$. The origin of this difference seems to be related to UV slope. Within the context of our model, $\beta$ is the strongest indicator of $f_{\rm esc}$. We re-emphasize that this seemingly also holds true for low-redshift ``analogs'' \citep{Chisholm_2022}. The $\beta$ values between the assumed high and low $f_{\rm esc}$ galaxies in \cite{Naidu2022} are formally consistent within the scatter, with a slight preference for the higher alleged $f_{\rm esc}$ sample to be more blue. Using the stacks they provide, we have estimated $f_{\rm esc}$ for their two samples and found values of 4\% and 0.4\% for the stacks with low peak velocity separation and high line-centre flux and high peak velocity separation and low line-centre flux, respectively. Applying the model from \citep{Chisholm_2022}, which only depends on $\beta$, we derive escape fraction values of 5\% and 3\%, respectively. The difference in $f_{\rm esc}$ in our model is primarily driven by the ${\rm E(B-V)}$ difference between the two samples. 

Our result does not indicate that bright Ly$\alpha$ emitters are insignificant for reionization. Despite their estimated lower escape fractions, their intrinsic production of ionizing photons is very high, and if 5\% of the ionizing photons leak, this may represent an important contribution to the emissivity budget. For the galaxies considered in \cite{Matthee2022} to dominate reionization one would need to increase the assumed $\xi_{\rm ion}$ to reconcile the lower escape fractions. Future JWST observations will undoubtedly provide new constraints on both the $f_{\rm esc}$ of Ly$\alpha$ emitters and $\xi_{\rm ion}$.

\subsection{Comparison with Other Simulations}

SPHINX$^{20}$ has not been the only attempt to study escape fractions in a cosmological context. Numerous works have been carried out studying how ionizing photons leak out of galaxies \citep[e.g.][]{Xu_2016,Barrow2020,Trebitsch_2021,Rosdahl_2022,Hassan_2022}. Our approach however has focused primarily on observational signatures. This is important, as we can directly study mock-observed galaxy properties, accounting for the anisotropic nature of LyC leakage. 

Simulations regularly show that $f_{\rm esc}$ is sensitive to stellar age. For example, haloes with stellar populations younger than $\SI{5}{\mega\year}$ in the FIBY simulations have higher escape fractions across a larger solid angle \citep{Paardekooper2015}. FIRE-2 find a lag between the timing of a star burst and an increase in $f_{\rm esc}$, due to the time needed for feedback to clear channels \citep{Ma_2020}, similar to what we find in SPHINX$^{20}$. They also found a telltale geometry when $f_{\rm esc}$ is high. Star-forming regions are surrounded by an accelerated, dense gas shell. Within this shell, young stars with ages $3-\SI{10}{\mega\year}$ are able to ionize low column-density channels through which radiation can leak. \cite{Kimm_2014} also find evidence for lags between star formation and high $f_{\rm esc}$. However, in their model, the lag was $\SI{10}{\mega\year}$. This highlights the sensitivity of this exact time delay to the SN model being used. In \cite{Kimm_2014}, star particles undergo SNe exactly $\SI{10}{\mega\year}$ after their birth, following \cite{Schaller_1992}. In contrast, the feedback recipe followed in SPHINX$^{20}$ includes a number of staggered SNe more accurately representing a single stellar population \citep{Kimm_2015}. Specifically, these begin as early as $\sim\SI{3}{\mega\year}$.

The need for ISM disruption has also been explored at great length. \cite{Ma_2020} find that star particles in galaxies with high escape fractions tend to be situated in regions with low column densities out to the virial radius. This is also corroborated by the findings of \cite{Trebitsch_2017} (see Figure 14). Moreover, \cite{Paardekooper2015} show that the neutral gas column density within 10~pc of a source is the defining quantity of escape fractions. \cite{Kimm_2014} agree, finding that galaxies with the lowest escape fractions tend to have the highest optical depths out to 100~pc and that the location of feedback is of vital importance. Particularly, the inclusion of runaway OB stars increases average escape fractions, due to the fact that these stars tend to move to lower density regions where the efficiency of feedback is greater. In our model, local gas density is included in our $\zeta_{\rm ISM}$ parameter. Because we have weighted the gas density by the [O~{\small II}]~$\lambda\lambda$3727 luminosity, we specifically pick out the densities in star-forming regions. The anti-correlation we find between $\zeta_{\rm ISM}$ and $f_{\rm esc}$ in SPHINX$^{20}$ is in agreement with these previous models.

\section{Caveats}
\label{sec:caveat}

Like all numerical simulations, SPHINX$^{20}$ employs a series of subgrid models for star formation, feedback, and ISM processes that could impact our results. For example, SPHINX$^{20}$ samples a distribution of SN time scales rather than assuming a fixed value which is why we find a critical time scale of 3.5~Myr for LyC leakage to begin. While assuming a delay time distribution is likely more realistic than a fixed value, a different delay time distribution would undoubtedly change which galaxies in SPHINX$^{20}$ are leakers, shifting the exact stellar age dependence stated. Likewise, our model for star formation assumes a variable efficiency of conversion from gas to stars. Changing this value will impact the clustering of stars and SN as well as the gas densities near young star particles. This will simultaneously affect the intrinsic emission line luminosities, the observed values (through a changing amount of dust as it is tied to the gas), and the efficiency of SN and radiative feedback. 

Since SPHINX$^{20}$ does not follow the formation and distribution of dust, we have employed an effective model where the dust-to-metal ratio is fixed and dust primarily tracks neutral gas \citep{Laursen_2009}. In contrast, observations show that the dust-to-gas mass ratio decreases following a power-law as a function of metallicity \citep{Remy-Ruyer_2014}. While the dust is often a sub-dominant contribution to the optical depth to ionizing photons at these redshifts \citep[e.g.][]{Katz_2022b} and therefore does not impact our escape fractions, the dust model does affect emission line luminosities and UV slopes. This motivates the study of IR lines as a probe of $f_{\rm esc}$ \citep[e.g.][]{Katz_2020,Ramambason2022} as they are significantly less sensitive to dust content. 

In this work, we have employed the BPASS SED which crucially extends the period over which ionizing photons are released due to binary interactions, compared to other SEDs. Similarly, one could change the model for Wolf-Rayet stars or include X-ray binaries which would similarly impact emission line fluxes. Our models also assume that metal abundance ratios match that of solar, whereas observations demonstrate that they likely vary as a function of metallicity. This will impact gas cooling and the state of the ISM as well as the emission line luminosities. For this reason, we have only worked with oxygen emission lines and their ratios which are likely well captured by our assumptions.

While currently the state-of-the-art for full-box reionization simulations, SPHINX$^{20}$ remains subject to limited spatial and mass resolution. This most importantly manifests in our inability to always fully resolve the Stromgren spheres of star particles. We have attempted to remedy this by using complex post-processing methods. However, the gas cells are still limited to have a fixed density on $\sim10$~pc scales. Subgrid density structure will impact emission line fluxes as well as dust attenuation and potentially change the impact of pre-SN feedback.

SPHINX$^{20}$ does not resolve the ISM to the same degree as simulations such as those in \cite{Kimm2019,Kimm2022}. In particular, \cite{Kimm2019} suggest that radiation feedback begins to disrupt the ISM of giant molecular clouds to the point of allowing LyC leakage at around $~\SI{2}{\mega\year}$, before SNe begin. In such a scenario, the window for effective LyC leakage would begin even earlier, inviting the need for further work with higher resolution simulations and better ISM and dust physics. However, it remains unclear whether gas outside the immediate molecular cloud prevents LyC photons from escaping into the IGM.

SPHINX$^{20}$ also neglects some potentially important physical processes such as stellar winds and cosmic rays. This could help lower the local densities around star particles and provide local metal enrichment. Both can impact emission line luminosities and the escape fraction. Likewise, we have neglected the nebular continuum. While properties such as ${\rm O_{32}}$ and $\Sigma_{\rm SFR}$ are unaffected, the nebular continuum can reduce observed equivalent widths and make $\beta$ appear redder. This is unimportant when $f_{\rm esc}\sim100\%$. Furthermore, the highest EW galaxies in our sample are non-leakers because they have not yet reached the SN time scale and it is these galaxies where the nebular continuum will be the most important.

Finally, SPHINX$^{20}$ does not include AGN and therefore captures neither the contribution of their hard radiation spectra nor of their feedback on the LyC escape fraction. As a result, it is important to note that the model derived in Section \ref{sec:GLM} should be used with caution on observations of galaxies with a confirmed AGN presence. In these cases, it is reasonable to expect that active feedback will clear ionized channels and therefore increase line-of-sight LyC escape fractions along the outflows, thus increasing the angle-averaged value for these galaxies. We leave such discussions to future work.

Despite these caveats, SPHINX$^{20}$ has been successful in reproducing numerous observations of the high-redshift Universe such as the UV luminosity function \citep{Rosdahl_2022} and Ly$\alpha$ luminosity function \citep{Garel2021}. The agreement we find with LzLCS is a promising sign that in many ways SPHINX$^{20}$ provides an adequate representation of the physics that leads to LyC leakage.

\section{Conclusions}
\label{sec:conclusion}

We have post-processed a sample of 1,412 star-forming galaxies in the SPHINX$^{20}$ cosmological radiation hydrodynamics simulation with {\small CLOUDY} and {\small RASCAS} to produce a diverse library of 14,120 simulated and dust-attenuated high-redshift galaxy spectra. These galaxies have been specifically selected to potentially be bright enough to be observable with JWST. Using this data-set, which represents part of SPHINX Public Data Release v1, we presented a new generalised framework for observational signatures of LyC leakage. Specifically, we argue that a good diagnostic to identify galaxies with significant LyC leakage should:

\begin{itemize}
    \item Track high sSFR;
    \item Select for stellar populations with ages $\SI{3.5}{\mega\year} \lesssim \langle \mathrm{Stellar\ Age}\rangle_{\rm LyC} \lesssim \SI{10}{\mega\year}$;
    \item Include a proxy diagnostic for neutral gas content as well as the state of the ISM.
\end{itemize}

This framework can successfully identify samples of galaxies that are highly enriched with LyC leakers. By applying our method to existing indirect $f_{\rm esc}$ diagnostics, we can predict the reasons why each diagnostic will be successful (or fail) and why. For example, we find that high O$_{32}$ is a necessary but insufficient criterion for high escape fractions, due to the fact that it traces high sSFR, but also selects for galaxies with dense, dusty ISMs with stellar populations that are too young to disrupt it. Observed UV slope, $\beta$, is empirically found to marginally satisfy two and strongly satisfy one criteria for $Z/Z_{\odot} > 0.01$. It is thus a relatively good diagnostic for the escape fraction. Similarly, ${\rm E(B-V)}$ is found to satisfy 2/3 criteria and thus traces $f_{\rm esc}$ reasonably well, albeit with significant scatter. In contrast, galaxy properties such as EW(H$\beta$), $\Sigma_{\rm SFR}$, $\Sigma_{\rm sSFR}$, ${\rm M_{UV}}$, sSFR, and ${\rm M_{*}}$ are all found to be poor indicators of $f_{\rm esc}$ if used in isolation as they satisfy one or none of our criteria. 

We can also satisfy all three criteria with multi-dimensional diagnostics. Selecting galaxies with $\log_{10}(\mathrm{H}\alpha/\mathrm{L}_{1500}) < 1.6$ while combining $\beta$ and ${\rm E(B-V)}$ (with Equation \ref{eq:observational_cut}) produces a sample of galaxies of which 67\% have $f_{\rm esc} > 5\%$, accounting for 62\% of all such galaxies in our data-set. Similarly, we have constructed a generalized linear model that utilizes spectral properties of galaxies to quantitatively predict $f_{\rm esc}$ (see Equation~\ref{eqn:fesc_predict}). Applying our model to high-redshift Ly$\alpha$ emitters observed with JWST, we find LyC escape fractions less than or equal to the observationally estimated Ly$\alpha$ escape fractions. Our results suggest that bright Ly$\alpha$ emitters tend to have LyC escape fractions $\lesssim 10\%$.

Though our framework for the physics of indirect estimators of LyC escape has been tested by a robust data-set of mock observations, we recognise that such mock data are dependent on the nebular emissivity and dust absorption models used, inviting future work on better-resolved cosmological simulations with more realistic sub-grid physics. The hardest of our three criteria to select for is the correct mean stellar population age, suggesting that this needs to be explored in the context of, for example, SED fitting. Nevertheless, our framework highlights the potential of existing and future JWST data to understand the physics of LyC escape and cosmological reionization.

\section*{Acknowledgements}
N.C. and H.K. thank Mengtao Tang for kindly sharing data on CEERS Ly$\alpha$ emitters. N.C. and H.K. also thank Jonathan Patterson for helpful support on Glamdring throughout the project.

N.C. acknowledges support from the Science and Technology Facilities Council (STFC) for a Ph.D. studentship.

This work used the DiRAC@Durham facility managed by the Institute for Computational Cosmology on behalf of the STFC DiRAC HPC Facility (www.dirac.ac.uk). The equipment was funded by BEIS capital funding via STFC capital grants ST/P002293/1, ST/R002371/1 and ST/S002502/1, Durham University and STFC operations grant ST/R000832/1. DiRAC is part of the National e-Infrastructure.This work was performed using the DiRAC Data Intensive service at Leicester, operated by the University of Leicester IT Services, which forms part of the STFC DiRAC HPC Facility (www.dirac.ac.uk). The equipment was funded by BEIS capital funding via STFC capital grants ST/K000373/1 and ST/R002363/1 and STFC DiRAC Operations grant ST/R001014/1. DiRAC is part of the National e-Infrastructure.

Computing time for the SPHINX project was provided by the Partnership for Advanced Computing in Europe (PRACE) as part of the ``First luminous objects and reionization with SPHINX (cont.)''  (2016153539, 2018184362, 2019215124) project. We thank Philipp Otte and Filipe Guimaraes for helpful support throughout the project and for the extra storage they provided us. We also thank GENCI for providing additional computing resources under GENCI grant A0070410560.  Resources for preparations, tests, and storage were also provided by the Common Computing Facility (CCF) of the LABEX Lyon Institute of Origins (ANR-10-LABX-0066) and PSMN (Pôle Scientifique de Modélisation Numérique) at ENS de Lyon.

\section*{Author Contributions}

The main roles of the authors were, using the CRediT (Contribution Roles Taxonomy) system\footnote{\url{https://authorservices.wiley.com/author-resources/Journal-Authors/open-access/credit.html}}:

\textbf{Nicholas Choustikov}: Conceptualization; Formal analysis; Writing - original draft; Methodology. \textbf{Harley Katz}: Conceptualization; Formal analysis; Writing - original draft; Methodology. \textbf{Aayush Saxena}: Conceptualization; Writing - review and editing. \textbf{Alex J. Cameron}: Conceptualization; Writing - review and editing. \textbf{Julien Devriendt}: Resources; Supervision; Writing - review and editing. \textbf{Adrianne Slyz}: Resources; Supervision; Writing - review and editing. \textbf{Joki Rosdahl}: Writing - review and editing. \textbf{Jeremy Blaizot}: Writing - review and editing. \textbf{Leo Michel-Dansac}: Writing - review and editing.
\section*{Data Availability}
The SPHINX$^{20}$ data used in this article is available as part of the SPHINX Public Data Release v1 \citep[SPDRv1,][]{spdrv1}.



\bibliographystyle{mnras}
\bibliography{References.bib}


\bsp	
\label{lastpage}
\end{document}